\documentclass[sigconf,screen]{acmart}

\usepackage[utf8]{inputenc}
\usepackage[normalem]{ulem}
\usepackage{tikz}
\usepackage{xspace}
\usepackage{booktabs}
\usepackage{multirow}
\usepackage{amsfonts}
\usepackage{bbding}
\usepackage{url}
\usepackage{listings}
\usepackage{longtable}
\usepackage{graphicx}
\usepackage{colortbl}
\usepackage[labelfont=bf]{caption}
\usepackage{ifpdf}
\usepackage{nowidow}
\usepackage{array}
\usepackage{multirow}

\makeatletter
\def\thickhline{%
  \noalign{\ifnum0=`}\fi\hrule \@height \thickarrayrulewidth \futurelet
   \reserved@a\@xthickhline}
\def\@xthickhline{\ifx\reserved@a\thickhline
               \vskip\doublerulesep
               \vskip-\thickarrayrulewidth
             \fi
      \ifnum0=`{\fi}}
\makeatother

\newlength{\thickarrayrulewidth}
\newif\iftwelve

\twelvefalse

\iftwelve\newcommand{\omittwelve}[1]{}
\else\newcommand{\omittwelve}[1]{#1}
\fi
\iftwelve\newcommand{\figexplainfont}{\footscriptsize}
\else\newcommand{\figexplainfont}{\normalsize}
\fi

\iftwelve\newcommand{\restablefont}{\footnotesize}
\else\newcommand{\restablefont}{\small}
\fi

\newcommand{\mycomment}[1]{}
\newcommand{\shepherdtwo}[1]{\shepherd{#1}}
\newcommand{\shepherd}[1]{#1}

\newcommand{\circled}[1]{\tikz[baseline=(char.base)]{
            \node[shape=circle,draw,inner sep=0.5pt] (char) {#1};}}
\newcommand{\ie}{\textit{i.e.,}}
\newcommand{\eg}{\textit{e.g.,}}
\newcommand{\etal}{\textit{et al.}}
\newcommand{\cf}{\textit{cf.}}
\renewcommand{\emph}[1]{\textit{\textbf{#1}}}
\newcommand{\code}[1]{\texttt{#1}}
\newcommand{\tool}{Privaros\xspace}
\newcommand{\myparagraph}[1]{\noindent\textbf{\textsf{#1}}}

\newcommand{\mycircled}[1]{\circled{\textsc{\textsf{#1}}}}
\newcommand{\bfcircled}[1]{\circled{\textbf{#1}}}
\newcommand{\policy}[1]{\textsc{#1}}
\newcommand{\topic}[1]{\textsf{#1}}
\newcommand{\app}[1]{\texttt{#1}}
\newcommand{\processlocally}{\policy{ProcessLocally}\xspace}
\newcommand{\blurexported}{\policy{BlurExportedImages}\xspace}
\newcommand{\usedronelanes}{\policy{UseDroneLanes}\xspace}

\newcommand{\negspace}[1]{\indent\vspace{#1}}
\newcommand{\mysection}[1]{\section{#1}}
\newcommand{\mycaption}[1]{\caption{\textbf{#1}}}
\newcommand{\sectref}[1]{Section~\ref{#1}}
\newcommand{\figref}[1]{Figure~\ref{#1}}

\iftwelve
    \newcommand{\mysubsubsection}[1]{\subsubsection{\textbf{#1}}}
\else
   \newcommand{\mysubsubsection}[1]{\subsubsection{\textbf{#1}}\indent\par\noindent}
\fi

\newcounter{myctr}
\newenvironment{mylist}{\begin{list}{\textbf{\bfcircled{\arabic{myctr}}}}
{\usecounter{myctr}
\setlength{\topsep}{1mm}\setlength{\itemsep}{0.5mm}
\setlength{\parsep}{0.5mm}
\setlength{\listparindent}{\parindent} 
\setlength{\itemindent}{0mm}\setlength{\partopsep}{0mm}
\setlength{\labelwidth}{-2mm}
\setlength{\leftmargin}{0mm}}}{\end{list}}

\newenvironment{mybullet}{\begin{list}{$\bullet$}
{\setlength{\topsep}{1mm}\setlength{\itemsep}{0.5mm}
\setlength{\parsep}{0.5mm}
\setlength{\listparindent}{\parindent} 
\setlength{\itemindent}{0mm}\setlength{\partopsep}{0mm}
\setlength{\labelwidth}{-2mm}
\setlength{\leftmargin}{0mm}}}{\end{list}}
\copyrightyear{2020}
\acmYear{2020}
\setcopyright{acmcopyright}
\acmConference[CCS '20] {2020 ACM SIGSAC Conference on Computer and Communications Security}{November 9--13}{2020}
\acmBooktitle{2020 ACM SIGSAC Conference on Computer and Communications Security (CCS '20), November 9--13, 2020}
\acmPrice{15.00}
\acmDOI{10.1145/3372297.3417858}
\acmISBN{978-1-4503-7089-9/20/11} 

\acmSubmissionID{fpc004}
\begin{document}
\iftwelve\else{}\fi

\title{\tool: A Framework for Privacy-Compliant Delivery Drones}

\author{Rakesh Rajan Beck}
\affiliation{\institution{Indian Institute of Science}}
\email{rakeshbeck@iisc.ac.in}

\author{Abhishek Vijeev}
\authornote{Rakesh Rajan Beck and Abhishek Vijeev contributed equally to this research.}
\affiliation{\institution{Indian Institute of Science}}
\email{abhishekvijeev@iisc.ac.in}

\author{Vinod Ganapathy}
\affiliation{\institution{Indian Institute of Science}}
\email{vg@iisc.ac.in}

\begin{abstract}


We present \tool, a framework to enforce privacy policies on drones.  \tool\ is designed for commercial delivery drones, such as the ones that will likely be used by Amazon Prime Air. Such drones visit various host airspaces, each of which may have different privacy requirements. \tool\ uses mandatory access control to enforce the policies of these hosts on guest delivery drones. \tool\ is tailored for ROS, a middleware popular in many drone  platforms. This paper presents the design and implementation of \tool's policy-enforcement mechanisms, describes how policies are specified, and shows that policy specification can be integrated with India's Digital Sky portal. Our evaluation shows that a drone running \tool\ can robustly enforce various privacy policies specified by hosts, and that its core mechanisms only marginally increase communication latency and power consumption.
\end{abstract}



\maketitle

\mysection{Introduction}
\label{sec:intro}

Over the past few years, there has been a rapid increase in the availability and ownership of end-user drones. Drones are now available for a few hundred dollars and widely used by hobbyists. Commercial operators such as Amazon are also planning to use fleets of drones for delivery. The US Federal Aviation Administration forecast report (2019-2039)~\cite{faaforecast} predicts over 1.39~million hobbyist and 853,000 commercial drones by 2023. Drones are also being put to effective use in the Covid19 pandemic. Law-enforcement agencies in various countries are using drones to make public-service announcements, and patrol locked-down areas for unauthorized social gatherings.

Despite the novel applications enabled by drones, the lack of tight regulations surrounding their use has led to a plethora of security and privacy problems. Incidents involving drones range from potential drone/aircraft collisions and near-misses~\cite{young:dailymail:2018}, drone-sightings causing airport closures~\cite{gatwick:shutdown}, to smuggling~\cite{bbc:smuggling:16} and assassination attempts~\cite{koettl:nytimes:2018}. While rogue drones cause such security and safety-related problems, benign drones, \eg~those that may be used for package delivery, also raise serious privacy concerns. Drones are equipped with a variety of sensors (cameras, GPS, Lidar, \textit{etc.}) for navigation. The sensors on board the drone can be used to capture pictures or video, map a sensitive location or a building. Prior studies have shown that people are indeed wary of their privacy being compromised by drones~\cite{perceptions:popets2016,spiders:chi2017,bystanders:chi2017}. Addressing the entire gamut of security and privacy problems posed by drones requires new regulations (\eg~to ensure that drones have an identity registered with the aviation authority), technology, and law enforcement (\eg~to detect~\cite{eshel13,birnbach:ndss:17,matthan:mobisys:2017,rozantsev:cvpr:2015,busset15,case:naecon08,vasquez:spie:08} and capture rogue drones). 

We present \textbf{\tool}, a framework that allows \textit{host airspaces} (\eg~a corporate or university campus, a city neighbourhood, or an apartment complex) to ensure that \textit{guest drones} entering them are compliant with privacy-policies determined by the hosts. For example, a host can specify a policy that requires any guest drone that enters its airspace to refrain from wirelessly transmitting or locally storing (\eg~in an on-board SD card) any images or video that it captures when within the host's airspace. \tool\ enhances the drone software stack with mechanisms that allow guest drones to enforce host-specified privacy policies and prove to the host that they are in compliance (via hardware-based attestations). 

We have designed \tool\ specifically with a focus on delivery drones. These drones are managed by fleet operators that have reputations and delivery contracts to protect and, by corollary, have no incentives to operate rogue drones. Thus, we can assume that such drones have an identity (\eg~a public key) that is registered with the aviation authority, are equipped with the \tool-enhanced software stack, and have associated trusted hardware that makes remote attestation possible. Making these assumptions allows us to focus on the key challenges in building the policy-compliance mechanisms for drones without rightaway having to consider other critical issues, \eg~on how to issue identity to drones and on how to deal with rogue drones. While central to an end-to-end treatment of privacy with drones, these issues involve developing regulations and evolving new law-enforcement methods that are outside the scope of this paper. Our work focuses on delivery drones because their usage model implies that the above assumptions hold. Our work also gels well with the policies being developed by various countries~\cite{brookings:bennett,brookings:mcneal,uk:caa,asg:drones,australia:oaic}, notably India's Digital Sky~\cite{digsky,digskyE1,digskyR1E1}, which provides guidelines for drone operators in India.

\tool\ models the problem of enforcing host-specified privacy policies as one of regulating how applications on the drone consume or communicate data received from sensors on the drone. \tool\ enforces these restrictions using mandatory access control. For the example policy discussed earlier, \tool\ can ensure that the video feed from the camera is available to image-processing/vision applications and to the navigation software, but cannot be sent to local storage or to the wireless network interface. 

\tool\ is built on top of the Robot Operating System~\cite{quigley2009ros, ros, ros2} (ROS version~2\footnote{Unless otherwise noted, uses of the term ``ROS'' in this paper reference ROS version~2~\cite{ros2}, which deviates significantly in design from ROS version~1~\cite{ros}.}), a popular middleware used by a number of drones (\eg~various models sold by DJI, 3DR, Parrot, Gaitech, Erle, BitCraze, Skybotix) and other robotics systems. A key reason for our choice to base \tool\ atop ROS was its rich support for applications written in a variety of languages, including Python and C++. ROS provides the abstractions to transparently execute a variety of applications on any drone hardware platform that runs ROS, and also interacts with the navigation control software. A vibrant application ecosystem has evolved around ROS and the market for platforms that use ROS is expected to grow to \$402.7 million by 2026~\cite{rosmarket}. \tool\ can therefore directly benefit drone operators that tap into the ROS application ecosystem.

ROS is built as a publish/subscribe system, in which ROS applications publish or subscribe to certain \textit{topics}. ROS simply acts as a matchmaker that pairs publishers and subscribers, following which the pair of applications communicate directly with each other over network sockets. As such, ROS does not incorporate any security mechanisms to regulate application communication. Thus, a malicious ROS application can easily snoop on or corrupt the communication between a pair of benign applications. Recognizing the need to prevent such attacks, the ROS community has developed Secure ROS (SROS)~\cite{white2019sros1,white2016sros}, a set of extensions that attempt to prevent such attacks.

In this paper, we show that the mechanisms of SROS alone do not suffice to robustly enforce security policies. In particular, while ROS applications typically communicate via the publish/subscribe mechanism, they can also communicate directly via other operating system (OS) abstractions, such as raw sockets, shared memory, pipes, and the file system. For example, a pair of applications can bypass the ROS-based publish/subscribe matchmaking mechanism, and directly establish socket connections for communication. While ROS has visibility into the publish/subscribe system and can reason about applications that initiate communication using this system, it cannot reason about low-level communication via OS abstractions.

\tool\ enhances the ROS software stack by adding the ability to enforce mandatory access control policies between ROS applications. It tightly integrates policy enforcement at the ROS layer with OS kernel-level modifications to enforce mandatory access-control policies. At the OS level, the mechanisms of \tool\ allow it to robustly enforce restrictions on applications that communicate directly via OS abstractions or bypass ROS. At the ROS level, \tool\ incorporates modifications that allow the OS mechanisms to be cognizant of ROS abstractions (\eg~topic names) used by applications, and suitably redirect communication via trusted applications, where required. \sectref{sec:design} elaborates on the mechanisms in \tool. 

We have tailored \tool's policy interface for delivery drone operations in India. To show that drones using \tool\ can readily be adopted once regulations are in place, we integrate \tool's policy specification interface with the front-end offered by India's Digital Sky portal~\cite{digsky}. This interface allows drone operators to specify the regions to which they intend to fly and obtain the permission to fly from India's Directorate General of Civil Aviation (DGCA). While the current intent of Digital Sky's interface is to prohibit drones from flying over so-called ``red-zones'' (\eg~military establishments or other sensitive areas), we show that the same interface can be used to upload the privacy policies of all the host airspaces that the drone will visit during its delivery run (\sectref{sec:implementation}). As a result, we hope that \tool\ can be readily adopted without invasive changes to a regulatory platform that is already in place. 

For our experiments, we ran the \tool-enhanced software stack on an NVidia Jetson TX2 board~\cite{TX2}. We chose this platform because its firmware can easily be reflashed with \tool\ (unlike off-the-shelf drones, which are often closed platforms), and also because it offers a programmable trusted-execution environment (TEE) based on the ARM TrustZone~\cite{arm2009security}. Commercially-available drones do not yet have the kind of hardware support to enable remote attestations by hosts. However, we note that such trusted hardware support has already been proposed as part of the regulations in Digital Sky~\cite{digskyR1E1}. Indian drone vendors will thus have to provide trusted hardware in the near future to sell and operate drones in India. Our evaluation (\sectref{sec:eval}) shows that \tool\ robustly enforces privacy policies. Furthermore, \tool's core mechanisms only introduce low runtime overheads in terms of communication latency and power consumption on the drone.

To summarize our contributions:
\begin{mybullet}
\item We motivate the problem of enforcing host-specified privacy policies on guest delivery drones and discuss the shortcomings of existing methods to enforce such policies;
\item We present \tool, a set of new mechanisms added to ROS and the underlying OS to enforce such policies;
\item We show how the policy specification for \tool\ can be integrated with the Digital Sky interface; and
\item We present a robustness and security evaluation of \tool\ on an NVidia Jetson TX2 board, and a performance evaluation showing that its overheads are low enough for practical use.
\end{mybullet}



\mysection{Background and Threat Model}
\label{sec:background}

We now present examples of the kinds of policies that we consider in this paper, background on the regulations that have already been proposed by Indian aviation authorities, and our threat model.

\iftwelve\else\indent\par\fi
\myparagraph{Example policies.} A host airspace may wish to impose a variety of policies on guest drones:
\begin{mylist}
\item \processlocally. Autonomously-navigated drones capture images or video of their surroundings. These images/video are processed on-board by a computer vision application to detect obstacles that the drone must fly around. A host may wish to ensure that the images/video captured by such a drone are only used by the computer vision application, which in turn communicates  this information only to the navigation board. In particular, the images/video must not be transmitted outside the drone via its network interfaces. They should also not be stored in the drone's filesystem for retrieval by the drone operator at a later point in time. 
\item \blurexported. A large majority of drones available today are controlled by a ground-based operator. These drones transmit a video feed from the drone to the operator (called the \textit{first person view}) who navigates the drone manually, often with visual line of sight. Alternatively, one could imagine an autonomously-navigated drone that transmits its video feed to a cloud-based server for obstacle detection, and obtains its navigation decisions from this server. For such drones, it is impractical for hosts to impose the \processlocally policy. Hosts may instead wish to ensure that the video feed exported from the drone is scrubbed to remove sensitive information. For example, the host may require the video feed to be processed by an on-board application (vetted by the host) that blurs peoples' faces and car registration plates that appear in the video feed (\eg~as is done in images published in Google Street View).
\item \usedronelanes. The host may require guest drones to fly only within designated \textit{drone lanes} to ensure safety and privacy. In a campus or university setting, campus security may identify drone lanes that are away from sensitive installations within the campus. Localities in a city may likewise identify drone lanes that border public spaces at a comfortable distance away from private homes.
\end{mylist}


\definecolor{LightGray}{gray}{0.95}

\begin{figure}[t!]
\centering
\iftwelve
\else
    \includegraphics[width=\linewidth]{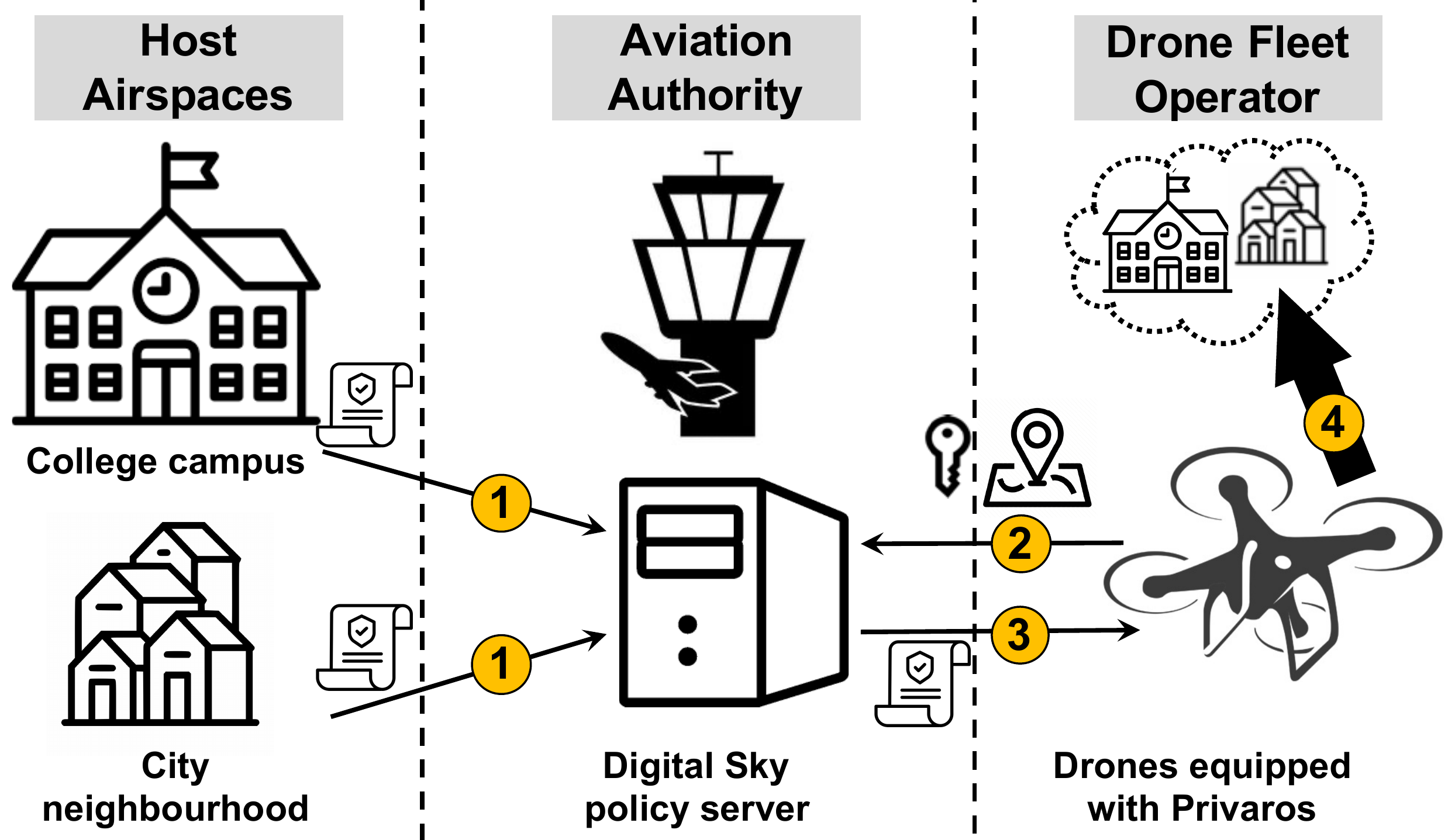}
\fi
{\figexplainfont
\textrm{\begin{tabular}{p{0.97\linewidth}}
\iftwelve\else\rowcolor{LightGray}\fi
\bfcircled{1}~Host airspaces, such as college/school campuses, office complexes/IT parks, and city neighbourhoods, identify their privacy policies and upload it to the aviation authority (Digital Sky in India);
\bfcircled{2}~A delivery drone operator starting up a drone sends information identifying the drone (\eg~its registered public key), and declares the flight path of the drone's delivery run;
\bfcircled{3}~The aviation authority server vets the flight plan (\eg~for NPNT compliance) and then sends the drone the privacy policies of the host airspaces along its flight path;
\bfcircled{4}~The drone then begins its delivery run. Any host airspace along the drone's flight path can challenge it to provide TEE-backed attestations, to verify the integrity of the drone's \tool-enabled software stack.
\end{tabular}}}
\iftwelve\else\negspace{-0.25cm}\fi
\mycaption{Deployment scenario for \tool-enabled drones.}
\iftwelve\else\negspace{-0.5cm}\fi
\label{figure:overallsetup}
\end{figure}

These policies have been discussed in prior work~\cite{vijeev:hotmobile:2019,alidrone2018icdcs}, but are by no means a comprehensive listing of policies that \tool\ can enforce.
Hosts may choose different combinations of these or other policies to enforce in their airspaces. The policies may also have to be tailored to the abilities of the guest drone, \eg~an autonomous drone can operate under \processlocally, but a semi-automated or manual drone may require \blurexported.

\shepherdtwo{These example policies cover the main concerns that have been raised in drone privacy laws proposed in several countries, \eg~the United States~\cite{brookings:bennett,brookings:mcneal}, United Kingdom~\cite{uk:caa,asg:drones}, Australia~\cite{australia:oaic}, and India~\cite{digskyE1,digskyR1E1}, where privacy of individuals and the integrity of the drone's flight path are the primary considerations. None of these proposals suggest a way to \textit{enforce} these policies beyond seeking an individual's permission before recording their picture or video.}

\iftwelve\else\indent\par\fi
\myparagraph{Digital Sky.} The Digital Sky platform is a set of regulations~\cite{digskyE1,digskyR1E1} and supporting computing infrastructure~\cite{digsky} that India's aviation authority (DGCA) is using to formulate its drone policy. A first set of regulations~\cite{digskyE1} was adopted on December 1, 2018 as part of the Civil Aviation Regulations, and the policy continues to evolve. Aimed primarily at drone operators, the Digital Sky portal~\cite{digsky} provides interfaces for authorized users (\eg~licensed commercial drone operators) to register the identity of their drones with DGCA, and obtain permission to fly before each delivery run. 

Digital Sky's policy is focused on ensuring safety and security. For example, Digital Sky allows operators to specify the geographic region over which they intend to fly using a visual map-based interface. This information is sent to a back-end server, which checks whether the region intersects any red-zones (\eg~sensitive military installations), which are no-fly zones. The proposed flight path may also intersect airspaces that impose altitude restrictions. Drones are not allowed to fly above a certain altitude as they approach an airport, for instance, and the permitted altitude reduces as the drone approaches the airport (thereby imposing a conical exclusion zone centred at the airport).

One of the centrepieces of Digital Sky's proposal to enforce these restrictions is \textit{No-permission No-takeoff (NPNT)}. With NPNT, the DGCA server would review the drone's flight path and issue a permission artifact, digitally-signed by the aviation authority. This permission artifact is sent to the drone, and the drone is permitted to fly only after validating the digitally-signed permission artifact. 

Finally, Digital Sky certifies drones at various levels~\cite[Page~39]{digskyR1E1}, based on the hardware capabilities of the drone. A level~0-certified drone stores cryptographic artifacts pertaining to its identity (\ie~its public/private key pair) completely within software. A level~1-certified drone must have a hardware-based TEE to store the drone's private keys, perform attestations, and perform NPNT validation and enforcement. Level~1-certified drones are robust to attacks on the drone's software stack because they offer a hardware-only TCB.

\tool\ aims to build upon the basic protections offered by Digital Sky by providing fine-grained policy enforcement within the drone. \tool\ allows enforcement of policies such as \processlocally, \blurexported\ and \usedronelanes\ that are beyond the current scope of Digital Sky.
\figref{figure:overallsetup} depicts how we envision \tool-enabled drones to be deployed, and how policies will be communicated to the drones. \sectref{sec:implementation} presents our deployment vision in more detail.

\iftwelve\else\indent\par\fi
\myparagraph{Threat model.} 
\tool\ is tailored for delivery drones. E-commerce companies considering drone-based delivery (\eg~Amazon, Flipkart) will likely use decentralized models akin to those used in ground-based delivery, wherein procurement and operation of delivery vehicles is outsourced to delivery-service providers (DSPs)~\cite{dsp:amazon,dsp:amazon:cbnc,dsp:flipkart}. While it is reasonable to assume that e-commerce companies are trusted and have no overtly-malicious intentions (because they have reputations to protect), host airspaces may not trust DSPs. In particular, the e-commerce company may prescribe a \tool-enhanced software stack for use on a DSP's delivery drones. However, the host airspace cannot trust that the DSP's guest drone is indeed executing that software stack. For example, the drone may have been compromised by a rogue DSP employee who covertly reflashes the drone's software or installs malware on it.

We therefore require guest drones to attest their software stack to host airspaces. Our trusted-computing base (TCB) consists of the guest drone's hardware TEE and its OS, enhanced with the mechanisms of \tool. The goal of attesting the drone is to ensure that the TCB in the guest drone is untampered. Applications running on top of \tool\ are not trusted and could be malicious in intent. \tool\ also introduces modifications to ROS (\sectref{sec:implementation:rosmodif}). However, ROS consists of library modules that are linked against applications. Since we do not trust applications, the modifications that \tool\ introduces in ROS are not part of the TCB.

To enable attestations, we assume that each drone is equipped with a hardware TEE (\ie~a level~1-certified drone in Digital Sky terminology) that stores the drone's private key. The drone's public key serves as its identity to hosts and the aviation authority. The hardware TEE enables features like secure boot and attestations in response to challenges from hosts or the aviation authority. We have implemented \tool\ on a hardware platform that has a TEE based on ARM TrustZone~\cite{arm2009security}. However, \tool\ only requires a TEE to attest the software stack (in the standard way~\cite{tpm:sec2004}) and its design does not currently leverage many other features of the TrustZone (\eg~peripheral partitioning across worlds). It can therefore be adapted to any hardware TEE design that drone vendors evolve in response to Digital Sky regulations.

A single drone typically has multiple compute platforms. For example, a \textit{flight control board} (such as Pixhawk~\cite{pixhawk}) runs the autopilot software (such as Ardupilot~\cite{ardupilot} or PX4~\cite{px4autopilot}, running on top of a real-time operating system such as ChibiOS~\cite{chibios}) and interacts with a \textit{companion board} (that typically uses an ARM-based processor) that runs applications. We assume that attestations provided by the drone cover the software running on \textit{all} these compute platforms. This could be implemented with a single \textit{master} board (\ie~the registered flight module whose identity is provided to the aviation authority) obtaining local attestations from all \textit{slave} boards, and providing a consolidated attestation to an external entity such as a host. Digital Sky requires all master-slave communications to be encrypted with 128-bit symmetric keys, at minimum~\cite[Page~39]{digskyR1E1}. For now, we only obtain attestations from the companion board that executes \tool. This is primarily because flight control boards are not currently equipped with hardware TEEs, although they are likely to evolve to be equipped as such.

By corollary, our threat model excludes physical attacks that attempt to bypass the mechanisms of \tool. For example, a rogue employee could attempt to bypass \tool\ by clipping on a remote-controlled camera with in-built networking that is not connected as a sensor to the compute platform running \tool. To an extent, some of these attacks can be mitigated with regulatory compliance checks. For example, regulations may require the fleet operator to have procedures that perform an automated physical check of the drone before it flies out of the warehouse on its delivery run to ensure that there are no unauthorized peripherals on the drone.

\tool\ provides the ability to control how applications consume sensor data. However, it is well-known that mandatory access control (\eg~based on subject and object labels) is a not perfect mechanism. It cannot protect against applications attempting to communicate with each other via covert timing or storage channels. We exclude covert channel-based attacks from our threat model.

\mysection{Enforcement Mechanism}
\label{sec:design}

This section presents the details of the policy enforcement mechanism in \tool. \sectref{sec:implementation} will describe how policies are specified and communicated to the drone.

\subsection{ROS}
\label{sec:design:ros}

\tool\ enhances ROS and the underlying OS, and mediates the actions of all the applications running on the drone. As discussed in our threat model, drone applications are typically executed on a companion board. This is standard in all ROS-based platforms. The flight-control board and the sensors connected to it communicate with applications on the companion board using the MAVLink protocol~\cite{mavlink}. Applications receive and process data from drone sensors and can also communicate with each other. For example, the output of the camera can be processed by an image-processing pipeline to detect obstacles. The output from this pipeline may be processed by a navigation application that sends MAVLink control commands to the flight-control board. \tool\ aims to control inter-application communication based on the host's privacy policies.

\omittwelve{
\begin{figure}[t!]
\centering
\texttt{
\begin{tabular}{lll}
\thickhline
\rowcolor{LightGray}
\multicolumn{3}{c}
{\underline{\textbf{Message format for topic \topic{CamOutput}}}}\\\rowcolor{LightGray}
uint32  & height     & \# image height\\\rowcolor{LightGray}
uint32  & width      & \# image width\\\rowcolor{LightGray}
string  & encoding   & \# encoding of pixels\\\rowcolor{LightGray}
uint32  & step       & \# row length\\\rowcolor{LightGray}
uint8[] & data       & \# image matrix (step*height)\\
\thickhline
\end{tabular}}
\negspace{-0.25cm}
\mycaption{Example declaration of message format in ROS. To send messages, an application uses a topic name (say \topic{CamOutput}) to refer to the message stream, and publishes messages in this format under that topic name.}
\negspace{-0.5cm}
\label{figure:rostopiceg}
\end{figure}}

ROS primarily uses a publish/subscribe model to facilitate application communication. The publish/subscribe mechanism in ROS is built using the  Data Distribution Service (DDS)~\cite{DDSI-RTPS,DDS_doc}, an open middleware standard created for real-time and embedded systems. ROS enables asynchronous communication between applications while decoupling spatial and temporal concerns, \ie~applications don't need to know where other applications that they communicate with reside (they can even run on a different drone), and applications can exchange information even if they are not simultaneously running. Applications publish or subscribe to one or more \textit{topics}, identified by a topic name. Topics have associated types that specify how messages published under that topic must be parsed. \omittwelve{\figref{figure:rostopiceg} presents an example of how an application would specify a ROS topic (in this case, an image). The fields shown in the topic declaration are the various data members of messages that are published under that topic.}

\begin{figure}[t!]
\centering
\iftwelve
\else
    \includegraphics[width=\linewidth]{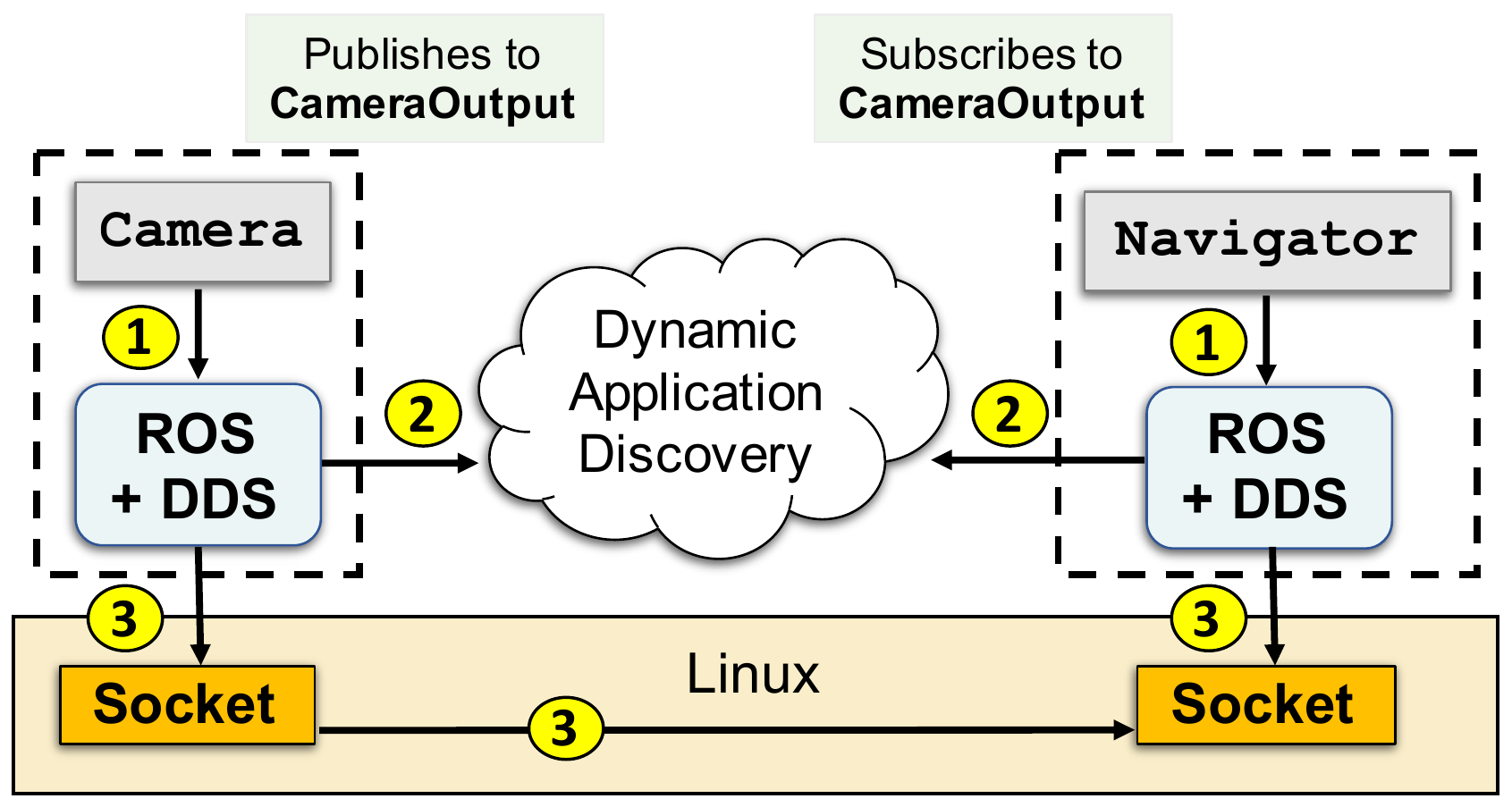}
\fi
{\figexplainfont
\textrm{\begin{tabular}{p{0.97\linewidth}}
\iftwelve\else\rowcolor{LightGray}\fi
\bfcircled{1}~Every application on ROS links against the library. The dotted lines show the process boundary. An application registers its topics via the ROS library;
\bfcircled{2}~A decentralized protocol discovers and identifies applications with matching topics;
\bfcircled{3}~ROS sets up socket communication via the underlying OS for the applications.
\end{tabular}}}
\iftwelve\else\negspace{-0.25cm}\fi
\mycaption{Publisher/subscriber communication in ROS.}
\label{figure:rosscomm}
\iftwelve\else\negspace{-0.5cm}\fi
\end{figure}

ROS uses the DDS protocol to match publishers with subscribers based on topic. At the application level, the abstraction presented is one of publishing messages to a bus, which are delivered to all subscribers of the bus. \figref{figure:rosscomm} illustrates how this abstraction is implemented. When ROS starts an application that subscribes to a topic, it checks which applications publish to that topic; DDS implements a decentralized protocol for application discovery. If it identifies a publisher, it sets up a network socket for the publisher and subscriber to directly communicate with each other. ROS is built to support distributed robotics platforms, so a publisher and subscriber application need not necessarily run on the same physical machine. However, if they do, ROS may choose to optimize their communication using shared memory instead of sockets. Applications that exclusively use the ROS API for communication remain oblivious to the means of communication (sockets, shared memory) that ROS uses to establish communication. The ROS library, which applications are linked with, transparently marshals and unmarshals data beneath the application layer, thereby exposing a simple publish/subscribe API at the application layer. 

We use a \textit{communication graph} to represent the flow of messages between applications on a ROS system. The nodes of this graph represent ROS applications while edges denote topic names. Note that each application can publish or subscribe to multiple topics. We present examples of communication graphs in \sectref{sec:design:privaros}.

\subsection{SROS and its Shortcomings}
\label{sec:design:sros}

In its most basic form, ROS does not offer security. Applications do not authenticate each other and messages between applications are exchanged in the clear. This leads to a number of attacks~\cite{dieber2016application,dieber2017security,rodriguez2018message,mcclean2013preliminary,jain2017dmtcp,wang2002security} that compromise message confidentiality (\eg~snooping on messages), data integrity (\eg~false data injection) and sender integrity (\eg~by impersonating an application). 

ROS does not impose  restrictions on the topics to which an application can publish or subscribe. This leads to situations where an application can publish a synthetic image feed with the same topic name as the real camera (say, \topic{CamOutput} is the topic name). Applications that subscribe to the \topic{CamOutput} topic will consume this image feed, possibly with dire consequences. For example, a malicious application can fool an obstacle-detection application that subscribes to \topic{CamOutput} by publishing an obstacle-free image feed, thereby causing the drone to crash into a building. Similarly, a malicious network-facing application can subscribe to \topic{CamOutput} and transmit the image feed to the attacker's server.

A number of prior papers have investigated these security shortcomings and have also proposed solutions~\cite{dieber2016application,dieber2017security,rodriguez2018message,mcclean2013preliminary}. The first version of ROS also had a centralized ROS master node, which was responsible for matchmaking. As a centralized entity, its failure could lead to denial-of-service attacks~\cite{jain2017dmtcp}; ROS version~2, which we use for \tool, eliminates the ROS master node, and instead uses the decentralized DDS protocol for communication setup. 

To address these concerns, the community has developed the Secure ROS extension (SROS)~\cite{white2019sros1,white2018procedurally}. SROS requires each node in the communication graph to be associated with an identity backed by a X.509 certificate, signed by a trusted third-party. SROS secures communication between nodes using TLS. It also allows application-writers to specify a manifest that lists the topics to which that application can publish or subscribe\omittwelve{ (\eg~see \figref{figure:manifesteg})}. The manifest is cryptographically bound to the application's identity and cannot be modified without regenerating associated the X.509 certificate. SROS thus ensures that an application cannot listen to or produce messages on topics that are not already part of its declared manifest.

\omittwelve{
\newcommand{\onetab}{\indent\hspace{0.5cm}}
\newcommand{\twotab}{\onetab\onetab}
\newcommand{\threetab}{\twotab\onetab}

\begin{figure}[t!]
\centering
\texttt{
\begin{tabular}{l}
\thickhline
\rowcolor{LightGray}
<permissions>\\\rowcolor{LightGray}
\onetab <grant name="/camera">\\\rowcolor{LightGray}
\onetab <subject\_name>CN=/camera</subject\_name> ...\\\rowcolor{LightGray}
\onetab <allow\_rule>\\\rowcolor{LightGray}
\twotab <publish>\\\rowcolor{LightGray}
\twotab <topics>\\\rowcolor{LightGray}
\threetab <topic>CamOutput</topic>\\\rowcolor{LightGray}
\threetab <topic>CameraStatus</topic>\\\rowcolor{LightGray}
\twotab </topics>\\\rowcolor{LightGray}
\twotab </publish>\\\rowcolor{LightGray}
\twotab <subscribe>\\\rowcolor{LightGray}
\twotab <topics>\\\rowcolor{LightGray}
\threetab <topic>Clock</topic>\\\rowcolor{LightGray}
\twotab </topics>\\\rowcolor{LightGray}
\twotab </subscribe>\\\rowcolor{LightGray}
\onetab </allow\_rule>\\\rowcolor{LightGray}
\onetab <default>DENY</default>\\\rowcolor{LightGray}
\onetab </grant>\\\rowcolor{LightGray}
</permissions>\\
\thickhline
\end{tabular}}
\negspace{-0.25cm}
\mycaption{Snippet of an application manifest in SROS.}
\label{figure:manifesteg}
\iftwelve\else\negspace{-0.5cm}\fi
\end{figure}}

These mechanisms prevent a number of basic attacks that are otherwise possible on a ROS system. But they are not perfect, and do not suffice to enforce policies end-to-end. We identify two fundamental, design-level shortcomings:
\begin{mylist}
\item \textit{Lack of end-to-end reasoning.} SROS restricts the list of topics to which an application can publish or subscribe via its manifest. However, when an application author specifies this list in the manifest, he does not know \textit{a priori} what other applications will execute on the drone platform. This lack of context-specific, end-to-end reasoning about the data produced or consumed by an application restricts our ability to enforce policies in arbitrary settings. For example, the \processlocally\ policy prevents any images published by the camera (under the topic \topic{CamOutput}) from being transmitted outside the drone. However, \blurexported\ does allow images to leave the drone as long as they are scrubbed by another application to blur any privacy-sensitive data in the images. The application author, who specifies the manifest, has no way to reason about all the contexts in which the application will execute. Without such reasoning about the application's end-to-end usage, the application author can at best produce a one-size-fits-all manifest that may poorly fit the situation in which the application is used.
\item \textit{Lack of control over lower-level abstractions.} SROS only imposes constraints on communication that goes via the ROS platform. Applications (both malicious and benign ones) can choose to bypass ROS entirely, and communicate directly with each other via network sockets, shared memory, the file system, or inter-process communication. Such communication happens directly via OS abstractions and therefore completely bypasses SROS enforcement.
\end{mylist}

In addition to these design-level shortcomings in SROS, we also identified some quirks in its implementation that could lead to unexpected attacks. First, SROS only allows application authors to specify restrictions in the manifest using \textit{topic names}. ROS version~2 allows an application to publish messages belonging to \textit{different types} under the \textit{same topic name}. For example, a camera application publishing under the topic \topic{CamOutput} could publish images under one type (say, \topic{CamOutput::ImageType}) and its status under another type (say, \topic{CamOutput::StatusType}). An application can choose to subscribe to messages of one or more of these types under the same topic. However, the type of data consumed by the application will not be evident in the manifest file, which only specifies the topic. For example, an application called \app{CameraStatus}\footnote{Throughout the paper, please note the font conventions used for \app{Applications}, \topic{Topics}, versus \policy{Policies}.} could periodically poll the camera's status by subscribing to \topic{CamOutput} and only read the data value published with type \topic{CamOutput::StatusType}. The fact that this application does not read the image feed from the camera is not evident from the manifest file.

The second quirk is that SROS internally uses the \textit{full path} of the application binary to identify the application at runtime. Using the path rather than the actual executable to determine identity makes the system vulnerable to attacks where the application binary is replaced with a malicious version. SROS will use the same manifest as the original application to determine the list of topics accessible to the malicious application.

Taken together, these quirks enable a malicious drone operator to engineer data leaks in certain situations. For instance, suppose the \app{CameraStatus} application is allowed to upload the camera's operational status to the network. A well-behaved \app{CameraStatus} application only reads data of type \topic{CamOutput::StatusType}, but not its image feed (of type \topic{CamOutput::ImageType}). A drone running such an application should therefore be acceptable to a host that wishes to enforce the \processlocally\ policy. However, if SROS were used for policy enforcement, a malicious drone operator could violate the \processlocally\ policy by replacing the \app{CameraStatus} application binary with a malicious version. The malicious application reads data of type \topic{CamOutput::ImageType} and leaks it over the network. SROS allows this attack because \mycircled{a}~it only uses the topic name in the manifest file to restrict the data channels accessible to the application; and \mycircled{b}~it only uses the path name of the binary and does not bind the executable to its identity.

We do not view these implementation-level quirks in SROS to be foundational. Indeed, there are easy workarounds: \eg~\mycircled{a}~modify ROS to include the type name with the topic in the manifest file (or match types at runtime); \mycircled{b}~rewrite applications to decouple different types of data into different topics; and \mycircled{c}~bind the application binary to its identity, possibly coupled with hardware TEE-based attestation of the binary to the host. However, the design-level shortcomings of SROS are the primary motivation for \tool.

\subsection{New Mechanisms in \tool}
\label{sec:design:privaros}

\tool\ enforces mandatory access control policies that regulate inter-application communication. \tool\ builds upon the basic facilities of SROS that assign identity to applications. It also leverages SROS to ensure that TLS is used for all inter-application network communication. However, it supplements the manifest-based access-control mechanism of SROS by:
\mycircled{a}~allowing end-to-end policy specifications; and
\mycircled{b}~enforcing policies within the OS.

\mysubsubsection{End-to-end Policy Specifications}In \tool, policies are specified in terms of permitted data flows between applications. Given a high-level policy such as \processlocally, \blurexported\ or \usedronelanes, the policy is compiled down to restrictions on inter-application communication (\sectref{sec:implementation} will discuss policy specification in more detail). Thus, rather than require an application writer to \textit{a priori} commit to specific topic restrictions, with \tool, restrictions to be imposed are identified based upon the environment in which the application will execute. Consider our three example policies, for instance:
\label{sec:design:privaros:end2end}

\begin{mybullet}

\item \processlocally. This policy is expressed using restrictions that prevent \mycircled{a}~any network-facing application from talking to the application that publishes the camera feed; and \mycircled{b}~preventing the camera application from writing to the file system mounted on the SD card. A navigation application may consume the output of the camera feed, but the policy would place the same restrictions on the navigation application (\ie~no network or file system communication) to prevent a leak of data from the navigation application.

\item \blurexported. This policy is compiled down to a restriction that all images from the camera must pass through a \textit{blurring application} before they are consumed by a network-facing application. The blurring application is entrusted with the task of identifying and blurring out faces, car number plates, and other sensitive data. 

\item \usedronelanes. This policy is compiled down to a restriction that the output of the GPS feed must pass through a \textit{trusted logger} that stores the GPS feed in tamper-proof storage, \eg~either in an audit log within the drone's hardware TEE, or in a trusted cloud server. Logs can later be analyzed to determine if the drone violated the drone lanes. 

\shepherd{The above implementation only allows passive enforcement of \usedronelanes, in which violations are detected \textit{post factum}. To actively enforce the policy, a trusted application would need to analyze the GPS feed and issue navigation commands to the flight-control board to keep the drone in the lane. We restricted ourselves to the passive enforcement variant because our experimental hardware platform~\cite{TX2} is not integrated with a flight-control board.}
\end{mybullet}

\begin{figure}[t!]
\small
\centering
\includegraphics[scale=0.35]{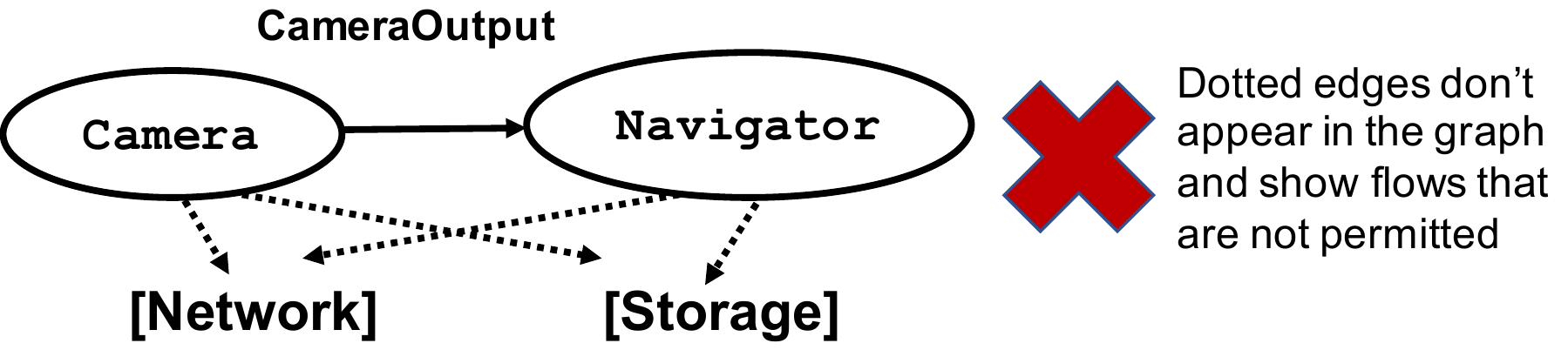}\\
\textbf{\mycircled{a}~Communication graph for \processlocally.}\\
\includegraphics[scale=0.34]{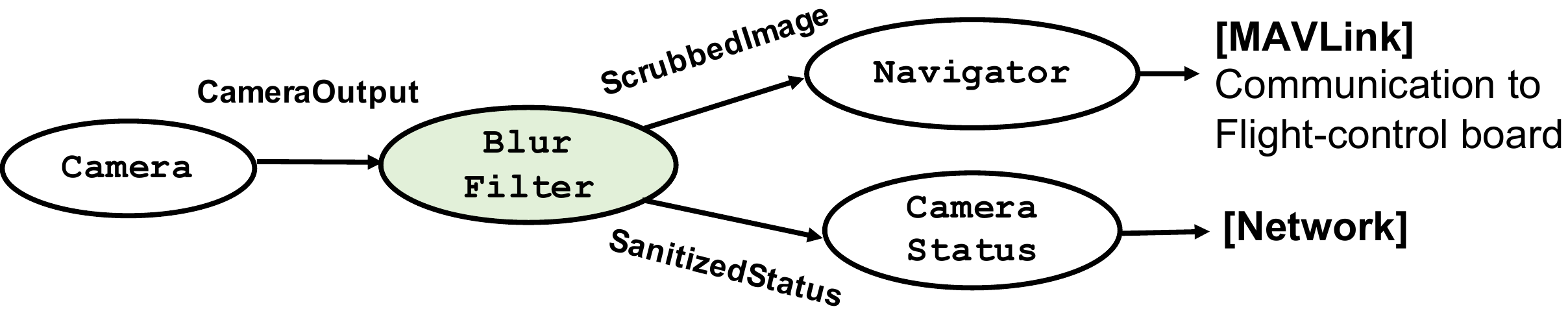}\\
\textbf{\mycircled{b}~Communication graph for \blurexported.}\\
\includegraphics[scale=0.34]{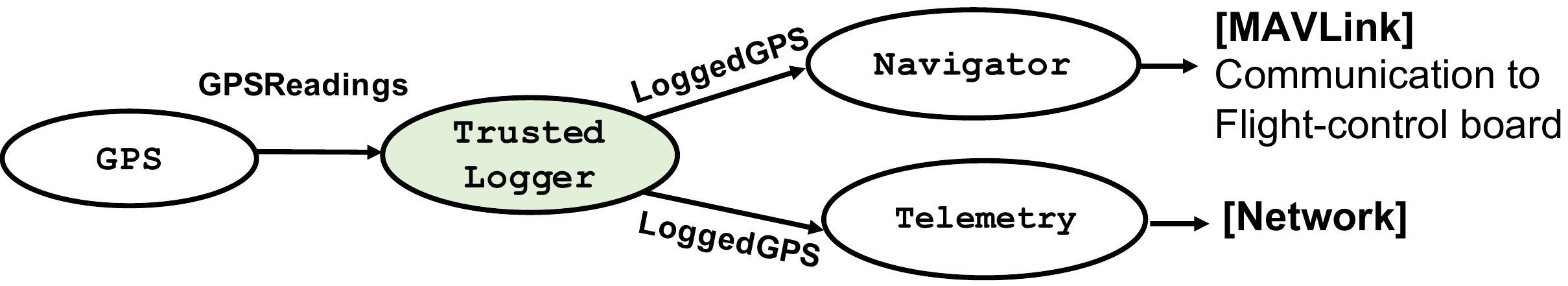}\\
\textbf{\mycircled{c}~Communication graph for \usedronelanes.}
\negspace{-0.25cm}
\mycaption{Illustrative communication graphs.}
\iftwelve\else\negspace{-0.5cm}\fi
\label{figure:commgrapheg-policies}
\end{figure}

\figref{figure:commgrapheg-policies} depicts the communication graph structure imposed by these restrictions. \tool\ relies on \textit{trusted applications} (\eg~the blurring application and trusted logger), shown as shaded ovals in \figref{figure:commgrapheg-policies}, to permit data flows that would otherwise be forbidden. These trusted applications serve a role similar to declassifiers/endorsers from the information-flow control literature~\cite{myers2000protecting} or transformation procedures from the Clark-Wilson security model~\cite{clarkwilson}.

The host that specifies the policy must also specify any trusted applications that may be needed to enforce the policy. These trusted applications may be drawn by the host from an app store-like portal. It is the host's responsibility to ensure that the trusted applications indeed meet their privacy requirements, \eg~that an app indeed identifies and blurs faces suitably. \tool\ confirms to the host that the trusted applications are executing on the drone (via hardware-based attestation, \sectref{sec:design:tee}), and ensures that data passes through these trusted applications before it reaches other downstream applications. \tool\ works at the granularity of processes, and does not track how data is processed within the applications to ensure that they perform their advertised functionality (\eg~blurring faces). 

\shepherd{Note that privacy laws of the future may require drones to enforce some of these policies by default. Even with such laws in place, we foresee \tool\ as being useful to hosts that wish to enforce customized policies. For example, a host may wish to specify and protect additional sensitive objects beyond those that are required to be blurred-out by law. Such hosts can use provide customized 
blurring applications so that \blurexported\ can identify and blur out sensitive objects of their choosing.}

\mysubsubsection{OS-level Enforcement}\tool\ restricts application-level communication within the OS (Linux, in our case) at the process level of granularity. Unlike SROS, \tool\ can therefore restrict application communication via pipes, the file system, shared memory, message queues, network- and UNIX-domain sockets. \tool\ validates the application binary at startup (using a digitally-signed hash of the binary) and enforces security policies on the corresponding process. As a result, \tool's enforcement binds the application's runtime identity to its process rather than the path name of its binary (\cf~\sectref{sec:design:sros}, the approach used by SROS).
\label{sec:design:privaros:os}

As is standard in many mandatory access control systems, \tool\ also uses labels to enforce policies~\cite{belllapadula,biba}. Each kernel object is tagged with a label; subjects (\ie~processes) also have labels. The labels of a subject and object determine whether the subject is allowed to access an object. An object's label can be changed by trusted endorsers or declassifers. This approach has been used in classical systems, \eg~Bell-LaPadula~\cite{belllapadula} and Biba~\cite{biba}, which use centralized labels determined by a system administrator. More recent approaches that implemented information-flow tracking in modern OS kernels have used expressive decentralized label systems, where applications decide the labels they assign to their data objects~\cite{taintdroid:osdi10,zeldovich2006making,krohn2007information,nadkarni2016sec}.

\tool\ adopts a simple label system that restricts data flow between subjects using mandatory access control. Policy rules are expressed at the process-level, and determine whether a process is allowed to create/read/write to sockets, shared memory, IPC, pipes or the file system. 

The approach of statically specifying which subjects can communicate with each other is somewhat more restrictive than the dynamic approach adopted by more expressive label systems (\eg~\cite{zeldovich2006making,krohn2007information,nadkarni2016sec,taintdroid:osdi10}). In these systems, the label associated with a data object encodes its dynamic security state, which stores the history of how it was processed (\eg~its taint status). In contrast, our policies are specified as static restrictions on subject (\ie~process) communication alone, and data labels do not feature in the policy specification. Thus, policy rules in \tool\ must be crafted carefully to keep track of the security state of an object. This difference has practical consequences in how a policy must be expressed. 

To illustrate the difference, consider enforcing the \blurexported\ policy on a drone that has a \app{Navigator} application which uses images of the camera to make local navigation decisions. However, suppose that \app{Navigator} also needs to occasionally transmit some of these images over the network to a cloud server for further analysis (\eg~if \app{Navigator}'s algorithms produce low confidence scores when identifying obstacles in those images). To enforce \blurexported, all images sent out over the network would have to be processed by a trusted \app{BlurFilter} application.

With a label system that tracks the dynamic state of data objects (\eg~\cite{nadkarni2016sec,zeldovich2006making,krohn2007information}), the label associated with the image will determine whether it has been processed by \app{BlurFilter}. There are no \textit{a priori} restrictions placed on \textit{when} \app{BlurFilter} should process the image. The only restriction is that the image should be processed by \app{BlurFilter} at some point during its lifetime before it is sent over the network.

In contrast, in \tool\ we encode \blurexported\ by placing restrictions on application communication. One way to express this would be using the communication graph in \figref{figure:commgrapheg-policies}\mycircled{b}, where we place the restriction that the camera application's output can only be consumed by \app{BlurFilter}, whose output in turn can be consumed by \app{Navigator} and other applications. 

\begin{figure}[t!]
\centering
\includegraphics[scale=0.35]{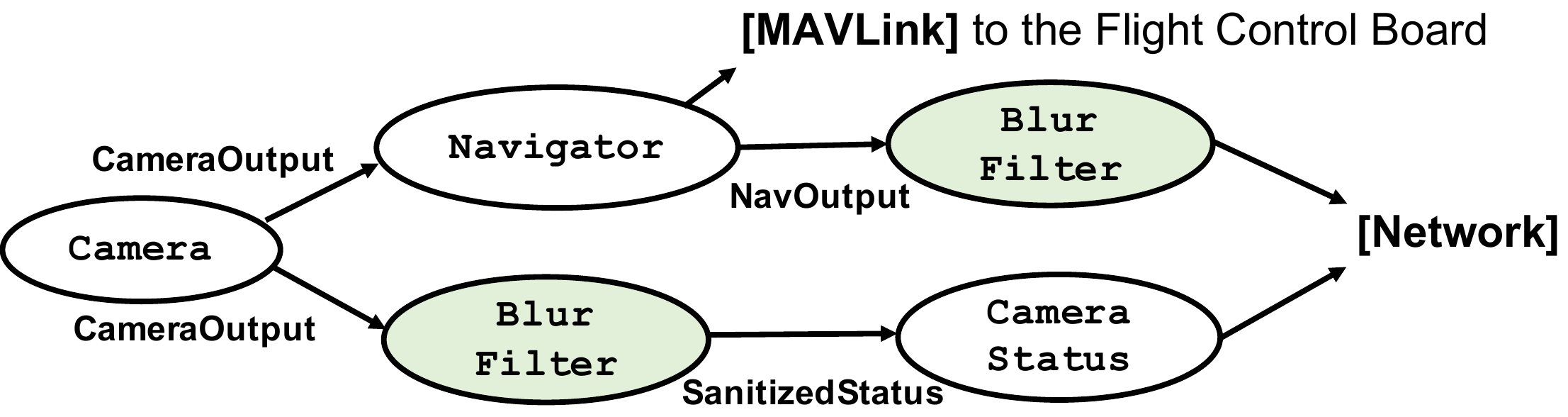}
\iftwelve\else\negspace{-0.25cm}\fi
\mycaption{Alternative communication graph to enforce \blurexported\ when \app{Navigator} needs a raw image feed and also has to be network-facing.}
\iftwelve\else\negspace{-0.5cm}\fi
\label{figure:commgrapheg-blurfilteralt}
\end{figure}

However, this is clearly not the only way to express this policy and may in fact be restrictive. For example, the \app{Navigator} application may require a high-fidelity image stream to make decisions, and the images processed by \app{BlurFilter} may not be of the desired quality. In this case, the desirable option would be to use the communication graph shown in \figref{figure:commgrapheg-blurfilteralt}. To realize this communication graph, \tool\ could either:
\mycircled{a}~run two instances of the \app{BlurFilter} application (as different processes), one for each node shown in the communication graph; or 
\mycircled{b}~only run one \app{BlurFilter} process, but modify the application to decouple the two logical flows. To process the first flow, \app{BlurFilter} would subscribe to \topic{CameraOutput} and publish that stream after processing to \topic{SanitizedStatus}. To process the second flow, it would subscribe to \topic{NavOutput} and publish scrubbed images to the network. \app{BlurFilter} must be configured carefully to segregate these flows. \tool\ must ensure that flows are directed to downstream applications correctly based on topic.

In \tool, we chose to express policies by statically restricting subject communication to keep the design of our enforcement mechanism simple. We found empirically that this approach works in the settings we considered. But it also means that policies must be crafted carefully to balance both the host's privacy requirements and the functionality of applications executing on the drone.

We have implemented \tool's enforcement mechanism using a kernel module. The kernel module hooks into the Linux Security Modules (LSM) framework~\cite{morris2002linux} to mediate kernel operations corresponding to various communication abstractions, and enforces the access control rules specified by the host. We base our implementation on AppArmor~\cite{apparmor}, so as to leverage their policy specification language and enforcement framework, which is quite mature and stable.  Applications may communicate through kernel abstractions such as pipes, files, network- and UNIX-domain sockets, shared memory, and message queues. \tool\ tracks such communication by attaching the label of the sending subject with the corresponding kernel abstraction. For example, we tag files with the identity of the process that created it (using \code{xattr}s, extended attributes provided on modern Linux file systems), and ensure that they can be read only by the same process or other processes as allowed by the policy. In the Linux kernel, most kernel abstractions provide extra fields to store such security state.

While \tool's in-kernel mechanisms are largely confined to the loadable kernel module, we did require some changes to the kernel itself in its networking subsystem. In particular, we found that when the LSM hook for the \code{sendmsg} system call is invoked, the recipient's information is not available from the \code{socket} data structure when the recipient is on the local host (\ie~the recipient's port is not yet bound). The kernel binds this information to the socket deeper down in the network stack. Therefore, we attach the sending process identifier with the \code{socket} data structure, and propagate this information as the \code{socket} descends down the network stack into the transport layer. When the packet is processed by the kernel for delivery to the recipient process, the identities of both the sender and receiver process are available, and \tool\ can decide whether the communication must be permitted.

As discussed earlier, ROS supports distributed robotics platforms, where the publisher and subscriber need not be on the same host. Thus, for instance, ROS can support fleets of drones where an application on one drone publishes data that can be consumed by applications running on other drones. Thus, network packets may leave the drone as well. One could consider a situation where a fleet of drones enforces \blurexported\ by running the \code{BlurFilter} application on just one drone (say, the fleet coordinator drone), and only allowing outbound network communications (\ie~out of the fleet) from that drone. In such cases, simply forbidding network packets containing the raw image feed from leaving a drone would be too restrictive. Instead, network packets must be allowed to the fleet coordinator, but not to other servers. \tool\ uses a whitelist of allowed domains (\eg~as done in Weir~\cite{nadkarni2016sec} and Hails~\cite{hails:osdi:12}) to allow such communication. The LSM hook for \code{sendmsg} determines whether the packet will leave the localhost, and if so, allows the communication only if the IP address of the destination (\eg~the IP address of the fleet coordinator) appears in a whitelist.

\mysubsubsection{Modifications to ROS}A key problem arises when \tool\ attempts to enforce policies such as \blurexported\ or \usedronelanes\ with off-the-shelf ROS applications. These policies require redirection of flows through trusted declassifiers before they can be consumed by downstream applications. However, the manifests of ROS applications will likely not allow redirection to happen easily. For example, consider a camera application's manifest that allows it to publish to the topic \topic{CamOutput} and a \app{Navigator} application whose manifest declares that it subscribes to \topic{CamOutput}. We cannot simply introduce a trusted \app{BlurFilter} application between the \app{Navigator} and the camera applications. \tool's OS-level mechanisms will permit the information flow from the camera process to the \app{BlurFilter} process and the output of the \app{BlurFilter} process to be consumed by \app{Navigator}, based on the policy. However, the ROS publish/subscribe system will not set up the flow because the topics do not match (\app{BlurFilter} publishes to \topic{ScrubbedImage}, to which \app{Navigator} has not subscribed).
\label{sec:implementation:rosmodif}

One way to address this problem is to generate manifest files for ROS applications based on the whitelisted flows in the communication graph. For example, \app{BlurFilter}'s manifest would declare that it subscribes to the topic \topic{CamOutput} and publishes to the topic \topic{ScrubbedImage}. In turn, \app{Navigator}'s manifest would allow it to subscribe to \topic{ScrubbedImage} (but not to \topic{CamOutput}). However, this approach may not be practical for off-the-shelf ROS applications whose manifests are part of their identity (\ie~X.509 certificates). The key difficulty is that fresh X.509 certificates have to be issued for each manifest configuration, which may not be feasible. 

\shepherd{\tool\ modifies ROS to allow flows to be transparently redirected between applications, as requested in the policy. In particular, it modifies the publish/subscribe system in ROS to:
\mycircled{a}~tear down an existing communication channel between a pair of applications;
\mycircled{b}~setup a new connection between applications, thereby allowing us to introduce a trusted declassifier; and
\mycircled{c}~assign a ROS topic and type to each newly-established connection.
Recall from \sectref{sec:design:sros} that application manifests only specifies the topic; the type is only available from the ROS runtime. \tool\ probes the publish/subscribe system to identify the type, and uses this information to annotate newly-added edges in the communication graph.}

Note that these modifications are required only to \textit{enable} communication between processes that is already permitted by the MAC-based enforcement of the OS. \tool\ relies solely on OS-level mechanisms to prevent applications from communicating. Thus, the modifications to ROS are \textit{not} part of our TCB. In particular, \tool\ allows a pair of applications to communicate only if allowed by both ROS and the OS's MAC-based policy enforcement.

Recall from \figref{figure:rosscomm} that applications use the ROS API via the ROS library that is linked into the process address space. The modifications discussed above are implemented within the ROS library and are transparent to ROS applications, which dynamically link against the ROS library on the drone platform. \tool's kernel-level mechanisms are also transparent to ROS applications.


\subsection{Role of the Hardware TEE}
\label{sec:design:tee}
As previously discussed, we use a TEE based on ARM TrustZone in our prototype implementation. Our prototype makes use of the TEE in the standard way for attestation~\cite{tpm:sec2004}. A TrustZone processor offers two \textit{worlds} of execution.  The \textit{normal world} executes untrusted applications and is typically the environment with which the end-user interacts. In our case, all drone applications, and Privaros (\ie~ROS and the OS-based mechanisms) run in the normal world. 

The \textit{secure world} manages the drone's private key,  implements remote attestation, and is therefore trusted and protected by secure boot. Its memory is isolated from the normal world. After booting securely, the secure world boots the normal world. It obtains and stores integrity measurements of the normal world boot process (\ie~a hash chain of software initialized during the boot sequence). These measurements can be used in remote attestations to convince a challenger (\eg~the aviation authority or any host airspace) that the normal world booted with an untampered TCB. The attestation report also includes a log of the applications started by \tool\ (as in TPM-based integrity measurement of applications executed over the system lifetime~\cite{tpm:sec2004}). Hosts can use this log to verify that any trusted declassifier applications that they entrust for policy enforcement in \tool\ are running on the drone.

\shepherd{Standard TEE-based attestation can detect attempts by a malicious DSP to install certain kinds of rootkits in the normal world. Rootkits that modify the normal world's kernel code or static data can be detected using integrity measurements at boot time. Although not currently implemented in our prototype, prior work has developed TEE-based methods to protect the normal world from advanced rookits, \eg~those that use direct kernel object manipulation. These methods have primarily been developed to offer real-time protection for kernel code~\cite{knox:ccs14,sprobes:2014} or obtain runtime snapshots of the normal world memory for analysis~\cite{trustdump:tifs:2015}. In addition, a CFI-protected~\cite{kcfi:eurosp:2016} normal world kernel (attested at boot-time using standard TEE-based integrity measurement) can provide real-time protection from various attacks directed against the kernel. We plan to integrate these methods in our prototype in future work.}

\mysection{Policy Interface}
\label{sec:implementation}

\myparagraph{Specifying and loading policies.}
\label{sec:implementation:policies}
Policies in \tool\ are specified using communication graphs. The graph identifies a whitelist of permitted flows between applications. Edges in the graph may be annotated with a topic name to denote the ROS topic that restricts the communication between that pair of applications to that topic alone. Edges may lack an annotation if applications are allowed to communicate outside the purview of ROS, \eg~using OS primitives.

Security administrators specify these policies by hand. However, a real-world drone may run dozens of ROS applications in addition to tens of daemons or other processes running natively on top of the underlying OS. For example, the communication graph on our experimental platform (\sectref{sec:eval}) has 29~nodes and 69~edges, even without any ROS applications running on it. Writing a comprehensive whitelist of allowed flows would therefore be time-consuming and might erroneously omit certain flows that prevent applications from working. We thus built a tool to extract communication graphs from a running drone (encapsulating all the flows between applications on that drone), which the security administrator can then use as a starting point and refine. We view this approach as being similar to the popular practice of using the \code{audit2allow} tool to write SELinux and SEAndroid policies. While we also fully acknowledge the usability concerns with  \code{audit2allow}, we view policy specification as an orthogonal problem that must be studied separately. Advanced policy analysis tools, such as those developed to configure SEAndroid policies~\cite{easeandroid,spoke:asiaccs:2017,combining:acsac:2017}, could be brought to bear as better alternatives to formulating policies.

Once a policy is written, it can be loaded into the drone for enforcement by \tool. We have built a user-level agent that identifies the process IDs of applications running on the drone, and translates the application names in the policy to the corresponding process IDs. \tool\ then applies the constraints imposed by this whitelist policy directly on the processes. The policy itself is expressed as a user-space file, but is serialized and loaded into the kernel via a user-agent (similar to the infrastructure provided by AppArmor, which \tool\ builds upon). The policy can be updated at any time by simply unloading the old policy and loading a new one, without restarting any applications. \tool\ thus transparently supports dynamic policy updates. This feature is important because dynamic policy updates may be required as the drone moves from one host airspace to another.

\iftwelve\else\indent\par\fi
\myparagraph{On integration with India's Digital Sky portal.}
\label{sec:implementation:digsky}
India's Digital Sky portal offers a Web-based service~\cite{digsky} via which drone operators indicate the proposed delivery route using a visual map-based interface. The Digital Sky server permits the delivery run if the route does not intersect any no-fly or other restricted zones.

We can extend the same interface for the setting that \tool\ considers. Each host specifies their privacy policies and geo-tags the policy with the coordinates of their airspace. The Digital Sky server stores a database of all registered hosts and their policies.

When a drone operator uses the Digital Sky server to mark the delivery route, the server identifies all host airspaces that the route intersects with (recall \figref{figure:overallsetup}). It then sends all the associated policies to the drone, where they are stored in the drone's local storage. The policies can be communicated to the drone using the same 
infrastructure (be it WiFi, 5G, or LTE) that the Digital Sky server uses to send NPNT approval certificates to the drone, prior to take-off. For the case of delivery drones, this step can happen at the warehouse from which the delivery run starts, where the availability of WiFi or wired network infrastructure can be assumed. Once the drone is airbrone, \tool\ continuously monitors the GPS coordinates of the drone, determines if it is entering a host airspace, and loads the corresponding policy from local storage for enforcement. It unloads the policy as it departs that host's airspace.

\begin{figure}[t!]
\centering
\iftwelve
\else
    \includegraphics[width=\linewidth]{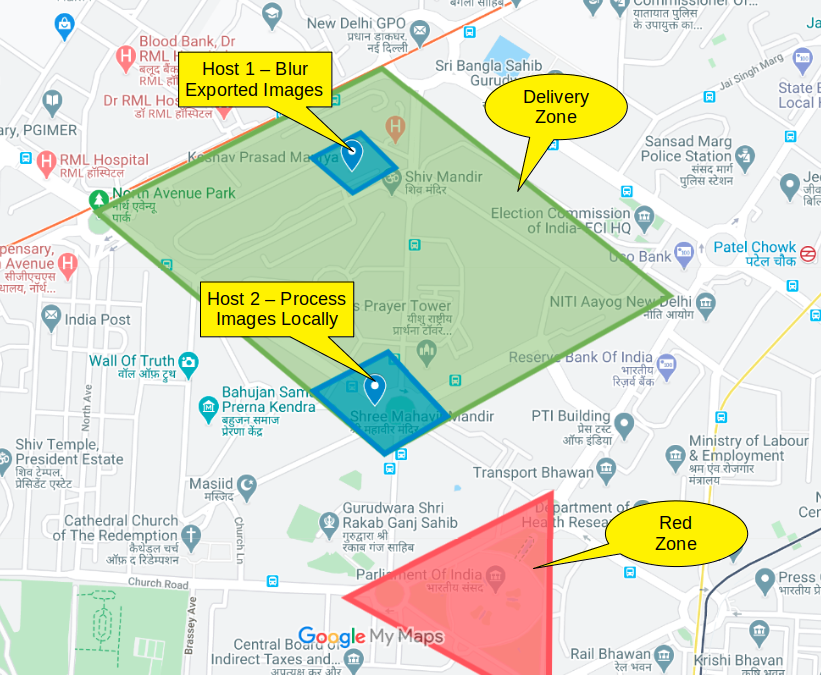}
\fi
\iftwelve\else\negspace{-0.25cm}\fi
\mycaption{
\iftwelve
\else
{Integrating policy specification in \tool\ with the Digital Sky interface.}
\fi}
\iftwelve\else\negspace{-0.65cm}\fi
\label{fig:digsky-privaros}
\end{figure}

We obtained the code of the Digital Sky Web server~\cite{digskygithub1,digskygithub2} and created a mock setup in our lab. \figref{fig:digsky-privaros} shows a screenshot of the Web server interface in which a drone operator has declared a drone's proposed delivery zone. It intersects two host airspaces, who have declared their privacy requirements. \figref{fig:digsky-privaros} also shows a red-zone (in this case, the Indian Parliament house in New Delhi; such sites would be identified by the aviation authority) that this drone's proposed delivery zone avoids.

Using the Digital Sky portal also has the benefit of simplifying the UI that a host would use for policy specification. Recall that \tool\ policies are specified as a communication graph of whitelisted flows. The key challenge in deploying this approach is that the communication graph must be customized for \textit{each policy} and \textit{each drone}. For example, to write the policy specification for \processlocally\ for \textit{a particular drone}, the host would have to \mycircled{a}~identify all network-facing applications on that drone; and \mycircled{b}~carefully create a communication graph in which the camera application never talks to a network-facing application. This exercise would have to be repeated for \blurexported, \usedronelanes\ and other policies of interest. And the whole exercise has to be repeated for every drone that potentially enters that host's airspace.

Digital Sky simplifies this exercise because it contains a database of all registered drones. This database could simply be extended to maintain a list of all applications installed on the drone, which Digital Sky can reliably obtain from the drone using the hardware TEE. We could pre-compute the communication graphs for various popular policy choices (\eg~\processlocally, \blurexported, \usedronelanes) in an offline fashion and store them in the database. From the host's perspective, the UI to specify policies can be simplified to a pull-down menu of common policy choices that they may wish to apply to their airspace. 
\iftwelve
\else
{When a drone expresses its intent to fly to the host's airspace, the Digital Sky server looks up the database to obtain the pre-computed communication graph corresponding to the combination of that drone and policy, and sends it to the drone.}
\fi

\mysection{Evaluation}
\label{sec:eval}


We implemented \tool\ on a system running Ubuntu 18.04 with Linux kernel version~4.9. We used ROS version~2 (Dashing Diademata) with eProsima FastRTPS version 1.8.2~\cite{eprosima_fastrtps,eprosima} as the underlying implementation of the DDS protocol. We enhanced it with the Secure ROS module available for ROS version~2 to enable TLS communication and to leverage the SROS application manifest infrastructure. We used AppArmor's user-space policy specification framework (version 2.13) to specify and download policies into the kernel for enforcement. Overall, we added or modified 402 lines of code in the ROS client library for C++ and 1651 in Linux to implement \tool\omittwelve{ (measured with \code{sloccount})}. We also modified 431 lines in the AppArmor user-space tool to parse policies, and added 213 lines of Python/bash code to support redirection of flows between ROS applications. 



We evaluated \tool\ on an Nvidia Jetson TX2~\cite{TX2} development kit, with a dual-core Denver 2 64-bit CPU and quad-core ARM A57 complex, 8 GB LPDDR4 memory and 32 GB eMMC flash storage. Our choice of Jetson was motivated by the fact that unlike most off-the-shelf drones, it is equipped with a hardware TEE (based on ARM TrustZone) and allows programmable access to both the secure world and the normal world. 
\shepherd{The specification of the Jetson board is similar in architecture to the companion boards of commercially-available drones. It also consists of 256 Nvidia CUDA cores, making it the companion board of choice for navigation software that makes extensive use of graphics processing units, \eg~those that use deep-learning based navigation.} We reflashed the normal world of this board with a \tool-enhanced software stack.

Our evaluation considers two questions: 
%
%
\mycircled{a}~How effective is \tool\ at enforcing policies, and how secure is it in comparison to SROS? (\sectref{sec:eval:robustness}); 
\mycircled{b}~What is the impact of \tool's mechanisms on latency, CPU utilization and power consumption, as evaluated with microbenchmarks? What is the impact of redirecting communication through trusted applications? (\sectref{sec:eval:performance}).
%

\subsection{Robustness of Policy Enforcement}
\label{sec:eval:robustness}
To showcase that \tool\ offers defense-in-depth, we built a malicious application that SROS cannot confine, and demonstrate the multiple layers \tool\ provides to confine this application.  Consider a \app{Camera} application that publishes to a topic called \topic{CameraOutput}. The application publishes two types of data under this topic: \mycircled{a}~the image feed from the camera under type \topic{CameraOutput::ImageType}, and \mycircled{b}~its status, under type \topic{CameraOutput::StatusType} (see \figref{figure:robustnesseval}\mycircled{a}). 

The primary goal of publishing \topic{CameraOutput::StatusType} is so that it can be consumed by \app{CameraStatus}, which is a benign application that subscribes to the topic \topic{CameraOutput} only to read the data published under the type \topic{CameraOutput::StatusType}. This application periodically uploads the camera's operating status to the drone operator's server that monitors the health of its fleet.

We examine various ways in which it is possible for an attacker to write a malicious application called \app{BadCameraStatus} that subscribes to the topic \topic{CameraOutput} but instead reads \topic{CameraOutput::ImageType} and transmits it over the network. The primary concern of a host would be to ensure the privacy of their image feed. We now examine how SROS and \tool\ compare in their ability to prevent the camera's output from being leaked either accidentally, or through malicious applications such as \app{BadCameraStatus}. We empirically validated each of the following attacks by implementing them and showing that \tool\ prevents them:

\begin{figure}[t!]
\centering
\small
\includegraphics[scale=0.35]{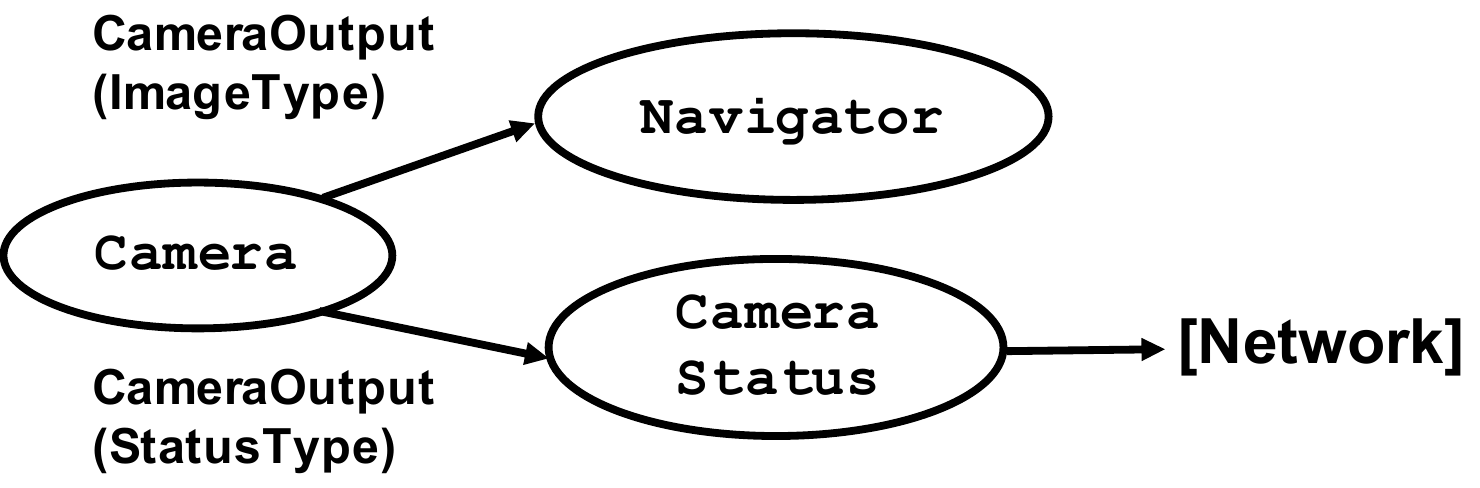}\\
\textbf{\mycircled{a}~Basic setup to be protected.}\\
\includegraphics[scale=0.33]{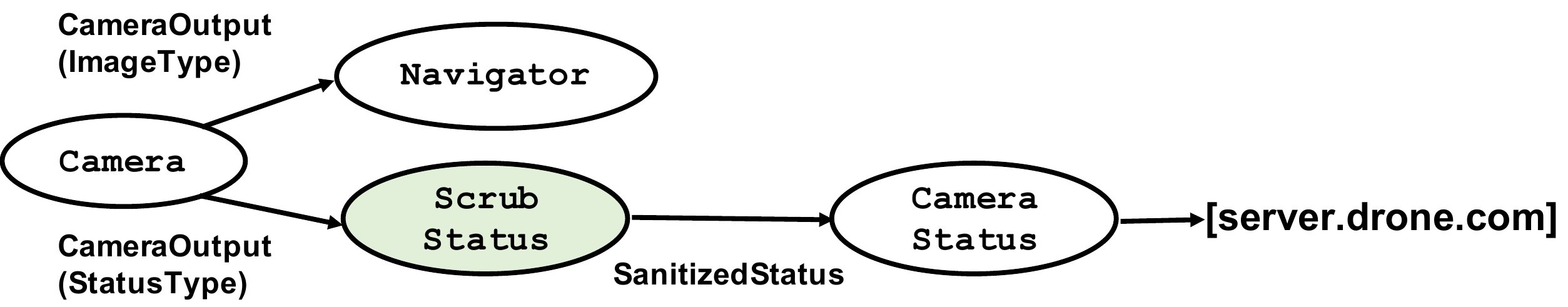}\\
\textbf{\mycircled{b}~Redirecting communication with \tool.}
\negspace{-0.25cm}
\iftwelve
\else\mycaption{Setup used in the experiments to demonstrate robustness of policy enforcement in \tool (\sectref{sec:eval:robustness}).}
\fi
\iftwelve\else\negspace{-0.5cm}\fi
\label{figure:robustnesseval}
\end{figure}

\begin{mybullet}
\item \textbf{Certificate checks.} The certificate checks in SROS can prevent an overt attempt at an attack, such as an attacker attempting to install \app{BadCameraStatus}. Under the assumption that such an application will not receive a valid certificate from a trusted authority, SROS certificate validation would fail, and SROS would not install the application. SROS would also prevent such an application (assuming it got installed) from subscribing to \topic{CameraOutput} if this is not declared in its manifest. For context, ROS (without SROS) would simply allow \app{BadCameraStatus} to be installed and allow it to subscribe to \topic{CameraOutput} and even publish messages to the same topic (\eg~a fake image feed).

However, the checks performed by SROS can easily be bypassed. An attacker (\eg~a malicious drone administrator) could replace the binary of the benign \app{CameraStatus} application with that of \app{BadCameraStatus} at the same file path (\cf~\sectref{sec:design:sros}). The attacker would launch this program using the same file path as the benign \app{CameraStatus} application, but it would perform the functionality intended by \app{BadCameraStatus}.
SROS checks X.509 certificates of apps, but does not associate the application's identity with their binary and instead only their full path name, and would therefore miss this attack. \tool\ prevents this attack because it checks the application binary's hash during certificate validation. 

\item \textbf{Redirection of app communication.}
To prevent accidental disclosure of the \app{Camera} application's image feed, a host could require that no network-facing application directly subscribe to \topic{CameraOutput}. It could instead require the camera's status to pass through a trusted application called \app{ScrubStatus}, which performs sanity-checks on the status feed. For example, \app{ScrubStatus} could ensure that the status feed only transmits a single byte. It could also rate-limit the flow (\eg~status updates allowed only once every 10 seconds), thereby mitigating the effects of any side-channels, via which an attacker attempts to leak images via the status feed\omittwelve{, byte-by-byte}.

One way to implement such enforcement in SROS would be to cleanly decouple the topics representing the image feed and the status feed by having the \app{Camera} application publish to two topics, \topic{ImageFeed} and \topic{StatusFeed} (because SROS only matches topics, and not types, as discussed in \sectref{sec:design:sros}). The \app{ScrubStatus} application could subscribe to \topic{StatusFeed}, but not \topic{ImageFeed}, and then publish the output to a topic \topic{SanitizedStatus} to which \app{CameraStatus} could subscribe and transmit over the network.

\tool\ can enforce this policy even if it is not easily possible to decouple the topics, \eg~because the \app{Camera} application code is not available or its manifest cannot be rewritten. With \tool, the trusted \app{ScrubStatus} application could still subscribe to \topic{CameraOutput}, but only read the \topic{CameraOutput::StatusFeed} type, and publish to \topic{SanitizedStatus} (see \figref{figure:robustnesseval}\mycircled{b}). Note that \tool's modifications to ROS (\sectref{sec:implementation:rosmodif}) are essential to allow \app{CameraStatus} to read the output of the \app{ScrubStatus} application. This is because the manifest of \app{CameraStatus} only allows it to subscribe to the topic \topic{CameraOutput} and not to \topic{SanitizedStatus}. However, \tool's modifications to ROS allow \app{CameraStatus} and \app{ScrubStatus} to communicate with each other.

\item \textbf{Direct communication via OS.}
The \app{BadCameraStatus} application could directly establish an inter-process channel (say, via UNIX domain sockets) to communicate with the \app{Camera} application, obtain images and send it over the network. SROS cannot mediate non-publish/subscribe communication and would allow this attack. The OS-level mechanisms of \tool\ prevent any communication between the processes unless allowed by the policy. Assuming the application redirection discussed above (through \app{ScrubStatus}), \tool\ can prevent any form of direct communication between \app{Camera} and \app{BadCameraStatus} (or even \app{CameraStatus}). All communication to network facing applications would have to go through the process that implements \app{ScrubStatus}.

\item \textbf{Whitelisting network domains.}
Finally, \app{CameraStatus} is a network-facing application. \tool\ uses whitelisting can ensure that the output of \app{CameraStatus} only goes to a particular IP address. SROS does not confine network communication this way.
\end{mybullet}

\subsection{Performance}
\label{sec:eval:performance}
We used microbenchmarks to measure the impact of \tool's core mechanisms on latency, CPU utilization, and power consumption. We used \app{PerformanceTest}~\cite{perf_test}, a DDS microbenchmark from Apex AI that is designed to evaluate the performance of publish/subscribe systems. \app{PerformanceTest} consists of a suite of workloads, each of which runs publishers and subscribers in different threads, and measures the latency involved in publishing/subscribing. \figref{figure:ubench} presents the details of the \app{PerformanceTest} workloads we used. We ran each workload configured to use one publisher and one subscriber, publishing at a rate of 10Hz for a duration of 10 seconds.

\begin{figure}[t!]
\centering
\restablefont
\begin{tabular}{l p{0.76\linewidth}}
\thickhline
\rowcolor{LightGray}
\textbf{\textsf{Workload}} 
    & \multicolumn{1}{c}{\textbf{\textsf{Type of data published/subscribed}}}\\
\hline
\app{Array}      & Simple byte array\\
\app{PointCloud} & 
    \iftwelve
\else
    {Collection of N-dimensional points (\eg~2D images produced by camera depth sensors)}
    \fi\\
\app{Struct}     & Structure holding a set of bytes (\eg~16 bytes in \app{Struct16})\\
\app{NavSat}     & Status of navigation satellite\\
\app{Range}      & Single range reading obtained from a range sensor\\
\thickhline
\end{tabular}
\negspace{-0.25cm}
\mycaption{Workloads from \app{PerformanceTest}~\cite{perf_test}.}
\iftwelve\else\negspace{-0.5cm}\fi
\label{figure:ubench}
\end{figure}

\app{PerformanceTest} reports the latency numbers for each workload. We measured the CPU utilization as the workload ran. To measure power consumption, we used the 3-channel INA3221 hardware power monitors on the Jetson TX2, which reports power draw of the board in milliwatts. \figref{figure:apexaiperftext} reports the results of our experiments. The baseline column reports the results of performing these experiments on a vanilla ROS/Linux setup with SROS enabled, and serves as the baseline. The \tool\ column reports the same numbers with the workloads running on \tool. As these numbers show, \tool\ imposes only a marginal increase in latency (under 10\% except when the latency numbers themselves are under 2ms) and power draw (under 5\% except in the case of \app{Struct32k}). Because drones are battery-powered, with current drones only providing an average flight time of about 20 minutes on a single charge, it is critical for \tool\ to be efficient with respect to power draw.  Finally, we also measured the performance impact imposed by \tool's hooks on individual kernel network subsystems using the \app{lmbench}~\cite{lmbench} benchmark. \figref{figure:lmbench} reports these results.

\begin{figure}[t!]
\restablefont
\begin{tabular}{l ccc}
\thickhline
\rowcolor{LightGray}
\textbf{\textsf{Workload}} 
    & \textbf{\textsf{Latency (ms)}} 
    & \textbf{\textsf{CPU (\%)}} 
    & \textbf{\textsf{Power (mW)}}\\
\hline\hline
\rowcolor{LightGray}
\multicolumn{4}{c}{\textbf{\textsf{\textit{Baseline}}}}\\
\hline
\app{Array1m} &
    16.255 &    
    6.728  &    
    2435.133
\\
\app{PointCloud1m} &
    16.160 &
    6.612 &
    2441.062
\\
\app{Struct32k} &
    6.494 &
    2.526 &
    2225.375
\\
\app{NavSat} &
    1.543 &
    1.381 &
    2349.353
\\
\app{Range} &
    1.433 &
    1.378 &
    2268.059
\\
\hline\hline
\rowcolor{LightGray}
\multicolumn{4}{c}{\textsf{\textbf{\textit{\tool}}}}\\
\hline
\app{Array1m} &
    17.225 (+5.9\%) &
    7.050 (+4.8\%) &
    2508.222 (+3.0\%)
\\
\app{PointCloud1m} &
    17.386 (+7.6\%) &
    7.141 (+8.0\%) &
    2437.294 (-0.2\%)
\\
\app{Struct32k} &
    7.109 (+9.5\%) &
    2.665 (+5.5\%) &
    2500.412 (+12.4\%)
\\
\app{NavSat} &
    1.922 (+24.6\%) &
    1.506 (+9.1\%) &
    2389.167 (+1.7\%)
\\
\app{Range} &
    1.928 (+34.5\%) &
    1.501 (+8.9\%) &
    2367.412 (+4.4\%)
\\
\thickhline
\end{tabular}
\iftwelve\else\negspace{-0.25cm}\fi
\mycaption{Microbenchmark performance.}
\label{figure:apexaiperftext}
\iftwelve\else\negspace{-0.25cm}\fi
\end{figure}

\mycomment{
\begin{figure}[t!]
\restablefont
\begin{tabular}{c|l ccc}
\thickhline
\rowcolor{LightGray}
\multicolumn{1}{c}{~}
    & \textbf{\textsf{Workload}} 
    & \textbf{\textsf{Latency (ms)}} 
    & \textbf{\textsf{CPU (\%)}} 
    & \textbf{\textsf{Power (mW)}}\\
\hline
\hline
\parbox[t]{3mm}{\multirow{5}{*}{\rotatebox[origin=l]{90}{\textbf{\textsf{\small Baseline}}}}}
& \app{Array1m} &
    16.255 &    
    6.728  &    
    2435.133
\\
& \app{PointCloud1m} &
    16.160 &
    6.612 &
    2441.062
\\
& \app{Struct32k} &
    6.494 &
    2.526 &
    2225.375
\\
& \app{NavSat} &
    1.543 &
    1.381 &
    2349.353
\\
& \app{Range} &
    1.433 &
    1.378 &
    2268.059
\\
\hline\hline
\parbox[t]{3mm}{\multirow{5}{*}{\rotatebox[origin=l]{90}{\textbf{\textsf{\textit{\small \tool}}}}}}
& \app{Array1m} &
    17.225 (+5.9\%) &
    7.050 (+4.8\%) &
    2508.222 (+3.0\%)
\\
& \app{PointCloud1m} &
    17.386 (+7.6\%) &
    7.141 (+8.0\%) &
    2437.294 (-0.2\%)
\\
& \app{Struct32k} &
    7.109 (+9.5\%) &
    2.665 (+5.5\%) &
    2500.412 (+12.4\%)
\\
& \app{NavSat} &
    1.922 (+24.6\%) &
    1.506 (+9.1\%) &
    2389.167 (+1.7\%)
\\
& \app{Range} &
    1.928 (+34.5\%) &
    1.501 (+8.9\%) &
    2367.412 (+4.4\%)
\\
\thickhline
\end{tabular}
\iftwelve\else\negspace{-0.25cm}\fi
\mycaption{Microbenchmark performance.}
\label{figure:apexaiperftext}
\iftwelve\else\negspace{-0.25cm}\fi
\end{figure}
}

\begin{figure}[t!]
\centering
\restablefont
\begin{tabular}{p{0.37\linewidth} c c }
\thickhline
\rowcolor{LightGray}
\textbf{\textsf{Workload}} 
    & \textbf{\textsf{Baseline}} 
    & \textbf{\textsf{\tool}}\\
\rowcolor{LightGray}
    & \textbf{\textsf{Latency ($\mu$s)}} 
    & \textbf{\textsf{Latency ($\mu$s)}}\\
\hline
Pipe &
    15.471 &
    15.640 (+1.093\%)
\\
UNIX domain sockets (TCP) &
    20.015 &
    23.188 (+15.9\%)
\\
UDP (localhost) &
    35.039 &
    35.374 (+1.0\%)
\\
TCP (localhost) &
    38.473 &
    38.764 (+0.8\%)
\\
UDP (RPC) &
    51.549 &
    52.335 (+1.5\%)
\\
TCP (RPC) &
    49.457 &
    49.977 (+1.1\%)
\\
\thickhline
\end{tabular}
\iftwelve\else\negspace{-0.25cm}\fi
\mycaption{Experiments using \app{lmbench}.}
\label{figure:lmbench}
\iftwelve\else\negspace{-0.5cm}\fi
\end{figure}

Finally, we studied the performance impact of redirecting data flow through a trusted application, as would be required for example to enforce \blurexported\ or \usedronelanes. Since these trusted applications are now part of the data-flow path, their presence will likely increase the latency of data delivery and overall power consumption. To illustrate the impact of trusted applications, we use the example of a \app{Camera} application whose images must pass through a trusted \app{BlurFilter} application before they are consumed by downstream applications. We measure the baseline performance, without application redirection, and two variants of the \app{BlurFilter} application: \mycircled{a}~a null-filter that simply redirects network flows but does not otherwise process the image (to measure the raw cost of redirection), and \mycircled{b}~a second one that is based on OpenCV, and blurs all image frames by 10\%. In this case, the application's processing logic itself performs non-trivial image-processing and consumes CPU and power. 

\figref{figure:redirection} reports the results of this experiment. The end-to-end latency of transmitting images from the camera to the network increases significantly when the \app{BlurFilter} application is introduced. The increase in latency is as expected, because of the additional user-space element involved in the outbound network path, and the associated transitions of the data packets between kernel-space and user-space. The end-to-end increase in power consumption remains under 10\% even with OpenCV-based blurring. 

\begin{figure}[t!]
\centering
\restablefont
\begin{tabular}{l c c}
\thickhline
\rowcolor{LightGray}
\textbf{\textsf{Scenario}} 
    & \textbf{\textsf{Latency (ms)}} 
    & \textbf{\textsf{Power (mW)}}\\
\hline
No redirection
    & 8.124 & 4749.400\\
\app{BlurFilter}/Null   
    & 17.509 (+115.5\%) & 4836.200 (+1.8\%)\\
\app{BlurFilter}/OpenCV
    & 21.511 (+164.8\%) & 5132.400 (+8.1\%)\\
\thickhline
\end{tabular}
\iftwelve\else\negspace{-0.25cm}\fi
\mycaption{Performance impact of flow redirection.}
\label{figure:redirection}
\iftwelve\else\negspace{-0.25cm}\fi
\end{figure}


Power consumption overheads will depend on the nature of processing involved in the trusted application. A real-world drone running dozens of applications will require many such trusted declassifiers, depending on the host policy to be enforced. Communication graphs must be carefully configured to minimize the number of distinct trusted elements required and their power consumption.

\mysection{Related Work}
\label{section:relwork}

\myparagraph{Drones and privacy.} To our knowledge, there has not been much prior work focusing on enforcing privacy policies in drones. In a prior paper~\cite{vijeev:hotmobile:2019}, we proposed the vision of restricted airpsaces for drones, in which host-specified policies would be enforced on guest drones. This paper builds upon that vision but makes significant additional contributions. In particular, this paper fully explores the challenges of building an enforcement framework, which was only sketched in our vision paper.  We learned that it was not possible to build an enforcement system on top of ROS alone, as 
outlined in our prior paper. We showed that even the primitives provided by SROS are insufficient to prevent a number of attacks  and that OS-level enforcement is central to ensuring robust policy enforcement. 
Finally, in this paper, we also showed how our policy-enforcement framework can be integrated with Digital Sky.

Nassi \etal~\cite{nassi:oakland:19} consider the problem of determining whether a drone's first-person view violates an individual's privacy. A first-person view projects the drone's camera feed as to a ground-based remote controller, operated by a human\omittwelve{(for purposes of navigation)}. The communication between the remote controller and the drone is encrypted. Nassi \etal\ develop a cryptanalysis technique by which an analyst with access to the encrypted first-person view feed can determine if the feed is focused on a particular object\omittwelve{(or person) of interest}. They apply physical perturbations to the object, \eg~by shining a light on it, at chosen points in time. The cryptanalysis determines if the encrypted feed is affected by these perturbations; if yes, they determine that the camera is focused on the object. They also develop techniques that spatially localize the offending drone by analyzing the first-person view feed.

\iftwelve\else\indent\par\fi
\myparagraph{Drones and security.} 
\iftwelve
\else
{In contrast to privacy, there is much prior work on security of drones. These range}
\fi
from using hardware TEEs to ensure that applications running on drones are able to securely access sensor data from its peripherals~\cite{liu2017protc} and to ensure that drones only fly along drone lanes~\cite{alidrone2018icdcs} (\ie~\usedronelanes), to 
investigating attacks against and protecting drones from common vulnerabilities, such as cleartext communication between drones, signal jamming and GPS spoofing~\cite{hartmann2013vulnerability,javaid2012cyber,wang2010cross,vattapparamban2016drones,pleban2014hacking,shepard2012evaluation,kerns2014unmanned,seo2016security}. \tool\ can benefit from techniques developed to defend against these attacks but has an orthogonal focus on enforcing host-specified privacy policies on drones.

The ROS community has also actively identified security vulnerabilities and attacks that stem from the unauthenticated, plaintext, publish/subscribe-based communication in ROS~\cite{dieber2017security,mcclean2013preliminary}. There have been proposals to use encrypted communication between applications~\cite{rodriguez2018message}, and to integrate TLS with the core libraries of ROS~\cite{dieber2017security,dieber2016application}. SROS~\cite{white2016sros,white2018procedurally,white2019sros1}, which is under active development, incorporates many of these ideas. As already discussed, \tool\ enhances the basic security features of SROS, eliminates some of its key shortcomings, and adds the ability to enforce privacy policies.

\iftwelve
\else
{While the above projects focus on securing drones from attacks, there is also work on detecting drones \ie~securing physical premises against unauthorized rogue drones.} 
\fi
These include methods to detect drones using their radar~\cite{eshel13} or radio-frequency signature~\cite{birnbach:ndss:17,matthan:mobisys:2017}, computer vision techniques to identify drones~\cite{rozantsev:cvpr:2015}, acoustic arrays that detect the sound of the drone's motors~\cite{busset15,case:naecon08}, and hybrid combinations of these techniques~\cite{vasquez:spie:08}. These techniques are undoubtedly important in formulating regulations to operate drones. However, they are orthogonal to \tool\ whose focus is on ensuring that authorized and legally-permitted drones conform to the privacy requirements of a host airspace.

\iftwelve\else\indent\par\fi
\myparagraph{Mandatory and context-based access control.} 
The idea of controlling the flow of information in computer systems can be traced back to some of the classic papers in computer security~\cite{belllapadula,biba,clarkwilson,mls:unix:1992}. SELinux~\cite{selinux}, SEAndroid~\cite{seandroid}, and related systems (\eg~\cite{flaskdroid:sec13,saint:acsac09}) have brought to bear some of these methods to modern OS settings. In these systems, subject and object labels are set by the system administrator, and the enforcement system applies label flow rules. 

In modern device-centric settings, some of these concepts have been adapted as context-based access control systems~\cite{flaskdroid:sec13,conxsense:asiaccs14,covington:acsac02,wdac:ccs14,deepdroid:ndss15,resp:mobisys16}, where the context in which the device is used (\eg~at home or in the workplace) determines the policies that must be enforced on the device. Some of these systems~\cite{flaskdroid:sec13,conxsense:asiaccs14,wdac:ccs14} also employ methods to actively infer the context in which the device is being used, and trigger the enforcement of the appropriate policy. \tool\ can also be viewed as a context-based access control system for delivery drones, and our focus in this paper has primarily been on building an enforcement mechanism integrated with ROS. Key contributions of this paper include:
\shepherd{\mycircled{a}~exploring the shortcomings of SROS;
\mycircled{b}~designing and implementing the cross-stack changes required for policy enforcement; and 
\mycircled{c}~redirecting flows through trusted applications for policy compliance.} 
\tool's policies are location-tagged, and the drone's GPS coordinates serve as the policy trigger to load access control policies. As discussed in \sectref{sec:implementation}, \tool\ allows dynamic loading/unloading of policies as a guest drone navigates between host airspaces. 

\iftwelve\else\indent\par\fi
\myparagraph{Information-flow control.}
Recent attempts to enforce information flow control on operating systems~\cite{zeldovich2006making,krohn2007information} and Android~\cite{taintdroid:osdi10,maxoid:eurosys:2015,nadkarni2016sec} have focused on \textit{decentralized} information-flow control (DIFC)~\cite{myers2000protecting}. DIFC systems differ from classic systems in that each application can specify its own labels, and the role of the system is to only use these labels to enforce certain rules on information flow. DIFC is particularly well-suited for settings where each application wants to control how \textit{its own data} is used by the rest of the system. In contrast, our setting requires us to apply host-specified privacy policies uniformly to all applications on the drone. \tool\ is therefore closer in spirit to the earlier work on using mandatory access control to regulate information flow~\cite{belllapadula,biba,clarkwilson}, SELinux~\cite{selinux} and SEAndroid~\cite{seandroid}. \tool\ adapts these concepts to a ROS-based platform and tightly integrate ROS-level and OS-level mechanisms.

\mysection{Conclusions}
\label{sec:conclusion}

\iftwelve
\else
{In this paper, we presented \tool, a framework that enforces privacy policies specified on guest drones visiting host airspaces for delivery runs. Our main conclusions are that:
\begin{mybullet}
\item
The problem of enforcing host-specified privacy policies on guest drones can be modeled as one of controlling the flow of data between applications executing on the drone;
\item
Existing mechanisms in ROS do not suffice to enforce these kinds of policies. Tight integration of ROS-level and OS-level mechanisms, as provided in \tool, are necessary for robust enforcement;
\item
Policy specification for \tool\ can be integrated with upcoming drone regulatory platforms such as Digital Sky;
\item
The core mechanisms of \tool\ impose low overheads on latency and power consumption. However, the host's policies may require the drone to execute trusted applications, which may themselves impose additional latency or consume additional power.
\end{mybullet}}
\fi

\omittwelve{
\indent\par
\myparagraph{Supplementary material.} 
Additional material related to this paper, including slides and code may be obtained from \url{http://www.csa.iisc.ac.in/~vg/papers/ccs2020}.
}

\iftwelve
\else
{\section*{Acknowledgments} 
We thank:
\mycircled{a}~John Johansen and Seth Arnold from the AppArmor development team for helping us resolve technical difficulties during the development of \tool; 
\mycircled{b}~Ruffin White for his insights on SROS, for illustrating the shortcomings of ROS1, and motivating us to develop \tool\ atop ROS2; 
\mycircled{c}~Earlence Fernandes for shepherding the paper and the reviewers from CCS'20 for their comments; 
\mycircled{d}~Robert Bosch Centre for Cyber-Physical Systems, IISc Bangalore, and a Ramanujan Fellowship from the Government for India, for generous grants that funded this work.
The last author dedicates this paper to the memory of his mentor and friend, Liviu Iftode of Rutgers University.}
\fi


\begin{center}
\footnotesize
URLs in references were last visited in August 2020.
\end{center}


\end{document}